%% file: paper.tex
\documentclass[twocolumn]{aastex6}

\usepackage{graphicx,lineno,svn-multi}
\usepackage{times}
\usepackage{mathptm}
\usepackage[nolist]{acronym}
\usepackage[utf8]{inputenc}

\hypersetup{
  citecolor=black,     % color of links to bibliography
}

\newcommand{\be}{\begin{equation}}
\newcommand{\ee}{\end{equation}}
\newcommand{\bsube}{\begin{subequations}}
\newcommand{\esube}{\end{subequations}}

\newcommand{\pycbc}{\texttt{PyCBC}}
\newcommand{\mbta}{\texttt{mbta}}
\newcommand{\gstlal}{\texttt{GstLAL}}

\usepackage{color}
\definecolor{cyan}{rgb}{0,0.9,0.9}
\definecolor{orange}{rgb}{0.9,0.5,0}
\definecolor{magenta}{rgb}{1,0,1}
\definecolor{purple}{rgb}{0.5,0.0,0.5}
\definecolor{teal}{rgb}{0.0,0.5,0.5}
\definecolor{gray}{rgb}{0.8242,0.8242,0.8242}

\input{./macros.tex}

\begin{document}

\title{Upper limits on the rates of binary neutron star and neutron-star--black-hole mergers from Advanced LIGO's first observing run} 
\input{LSC_Feb2016_Virgo_Feb2016-aas.tex}
\date{\today}

\begin{abstract}
We report here the non-detection of gravitational waves from the merger of binary neutron star systems and
neutron-star--black-hole systems during the first observing run of Advanced LIGO.
In particular we searched for gravitational wave signals from binary neutron
star systems with component masses $\in [1,3] M_{\odot}$ and component dimensionless
spins $< 0.05$. We also searched for neutron-star--black-hole systems with the same
neutron star parameters, black hole mass $\in [2,99] M_{\odot}$ and no restriction
on the black hole spin magnitude.
We assess the sensitivity of the two LIGO detectors to these systems, and 
find that they could have detected the merger of binary neutron star
systems with component mass distributions of $1.35\pm0.13 M_{\odot}$
at a volume-weighted average distance of \MainBNSRange, and for neutron-star--black-hole
systems with neutron star masses of $1.4M_\odot$ and black hole masses 
of at least $5M_\odot$, a volume-weighted average distance of at least \MainNSBHRangeFive.
From this we constrain with 90\% confidence the merger rate to be less than
\MainBNSULPyCBCHighSpin~Gpc$^{-3}$~yr$^{-1}$ for binary-neutron star systems
and less than \MainNSBHULPyCBCFiveIso~Gpc$^{-3}$~yr$^{-1}$ for
neutron-star--black-hole systems.
We discuss the astrophysical implications of these results, which we find to be in tension with only the
most optimistic predictions. However, we find that if no detection of neutron-star binary mergers is
made in the next two Advanced LIGO and Advanced Virgo observing runs we would place significant
constraints on the merger rates.
Finally, assuming a rate of $10^{+20}_{-7}$Gpc$^{-3}$yr$^{-1}$ short gamma ray bursts beamed towards the Earth
and assuming that all short gamma-ray bursts have binary-neutron-star (neutron-star--black-hole)
progenitors we can use our 90\% confidence rate upper limits to constrain the beaming angle of the gamma-ray
burst to be greater than \GRBBNSBeamingAngleConstraint\ (\GRBNSBHFiveBeamingAngleConstraint).

\end{abstract}

\maketitle

\input{acronyms}

\section{Introduction}

Between \OoneSTART\ and \OoneEND\, the two advanced \ac{LIGO} detectors conducted their \ac{O1}.
During \ac{O1}, two high-mass \ac{BBH} events
were identified with high confidence ($> 5 \sigma$): GW150914~\citep{Abbott:2016blz} and
GW151226~\citep{Abbott:2016nmj}. A third signal, LVT151012, was
also identified with \LVBLAHsignificance\ confidence~\citep{TheLIGOScientific:2016pea, TheLIGOScientific:2016qqj}
In all three cases the component masses are confidently constrained to be above the $3.2M_\odot$ upper mass limit of \acp{NS} set
by theoretical considerations~\citep{Rhoades:1974fn,TheLIGOScientific:2016wfe}.
The details of these observations, investigations about the properties
of the observed \ac{BBH} mergers, and the astrophysical implications are explored
in~\citep{TheLIGOScientific:2016wfe,Abbott:2016nhf,TheLIGOScientific:2016htt,TheLIGOScientific:2016src,TheLIGOScientific:2016pea, Abbott:2016izl}.

The search methods that successfully observed these \ac{BBH} mergers also target other types of compact
binary coalescences, specifically the inspiral and merger of \ac{BNS} systems and \ac{NSBH} systems. Such systems were considered
among the most promising candidates for an observation in \ac{O1}. For example, a simple calculation
prior to the start of O1 predicted 0.0005 - 4 detections of \ac{BNS}
signals during O1~\citep{Aasi:2013wya}.

In this paper we report on the search for \ac{BNS} and \ac{NSBH} mergers in \ac{O1}. We have
searched for \ac{BNS} systems with component masses $\in [1,3] M_{\odot}$, component dimensionless
spins $< 0.05$ and spin orientations aligned or anti-aligned with the orbital angular momentum.
We have searched for \ac{NSBH} systems with neutron star mass $\in [1,3] M_{\odot}$,
\ac{BH} mass $\in [2,99] M_{\odot}$ neutron star dimensionless spin magnitude $< 0.05$,
\ac{BH} dimensionless spin magnitude $<0.99$ and both spins
aligned or anti-aligned with the orbital angular momentum.
No observation of
either \ac{BNS} or \ac{NSBH} mergers was made in \ac{O1}. We explore the astrophysical implications
of this result, placing upper limits on the rates of such merger events in the
local Universe that
are roughly an order of magnitude smaller than those obtained with data from Initial \ac{LIGO}
and Initial Virgo~\citep{Abbott:2007kv,Acernese:2008zzf,Colaboration:2011np}.
We compare these updated rate limits to current predictions of \ac{BNS} and
\ac{NSBH} merger rates and explore how the non-detection of \ac{BNS} and \ac{NSBH} systems in \ac{O1} can be used
to explore possible constraints of the opening angle of the radiation cone of short \acp{GRB},
assuming that short \ac{GRB} progenitors are \ac{BNS} or \ac{NSBH} mergers.

The layout of this paper is as follows. In \S\,~\ref{sec:source_considerations} we describe
the motivation for our search parameter space. In \S\,~\ref{sec:search_description} we briefly
describe the search methodology, then describe the results of the search in \S\,~\ref{sec:search_results}.
We then discuss the constraints that can be placed on the rates of \ac{BNS} and \ac{NSBH} mergers in \S\,~\ref{sec:rates}
and the astrophysical implications of the rates in \S\,~\ref{sec:astrophys_interp}. Finally, we conclude
in \S\,~\ref{sec:conclusion}.

\section{Source considerations}
\label{sec:source_considerations}

There are currently thousands of known NSs, most detected as pulsars
\citep{pulsarcat,Manchester:2004bp}. Of these, $\sim70$ are found in binary
systems and allow estimates of the NS mass
\citep{nsmassespage,Lattimer:2012nd,Ozel:2016oaf}.
Published mass estimates range from $1.0\pm0.17\,{M_{\odot}}$ \citep{Falanga:2015mra} to
$2.74\pm0.21\,{{M_{\odot}}}$ \citep{Freire:2007jd} although there is some uncertainty
in some of these measurements.
Considering only precise mass measurements from these observations one can set a lower bound
on the maximum possible neutron star mass of $2.01\pm 0.04\,{M_{\odot}}$~\citep{Antoniadis:2013pzd} and theoretical considerations
set an upper bound on the maximum possible neutron star mass of $2.9$--$3.2\,{M_{\odot}}$ \citep{Rhoades:1974fn,Kalogera:1996ci}.
The standard formation scenario of core-collapse supernovae restricts the birth
masses of neutron stars to be above $1.1$--$1.6\,{M_{\odot}}$
\citep{Ozel:2012ax,Lattimer:2012nd,Kiziltan:2013oja}.

Eight candidate \ac{BNS} systems allow mass measurements for individual
components, giving a much narrower mass distribution~\citep{Lorimer:2008se}. Masses are
reported between $1.0\,{{M_{\odot}}}$ and 
$1.49\,{{M_{\odot}}}$~\citep{nsmassespage,Ozel:2016oaf}, and are consistent with an
underlying mass distribution of $(1.35 \pm
0.13)\,{{M_{\odot}}}$~\citep{Kiziltan:2010ct}. 
These observational measurements assume masses are greater than $0.9{{M_{\odot}}}$. 

The fastest spinning pulsar observed so far rotates with a frequency of 716\,Hz~\citep{Hessels:2006ze}. This
corresponds to a dimensionless spin $\chi = c | \mathbf{S} | / G m^2$ of roughly 0.4, where $m$ is the object's
mass and  $\mathbf{S}$ is the angular momentum.\footnote{Assuming a mass of $1.4{{M_\odot}}$ and
a moment of inertia $=J/\Omega$ of $1.5\times10^{45}$\,g\,cm$^2$; the exact moment of 
inertia is dependent on the unknown \ac{NS} equation-of-state~\citep{Lattimer:2012nd}.} Such
rapid rotation rates likely require the NS to have been spun up through mass-transfer
from its companion. The fastest spinning pulsar in a confirmed \ac{BNS} system has a
spin frequency of 44\,Hz~\citep{Kramer:2009zza}, implying that dimensionless
spins for NS in \ac{BNS} systems are $\leq 0.04$~\citep{Brown:2012qf}. However,
recycled NS can have larger spins, and the potential \ac{BNS} pulsar J1807-2500B
\citep{Lynch:2011aa} has
a spin of 4.19\,ms, giving a dimensionless spin of up to
$\sim0.2$.\footnote{Calculated with a pulsar mass of $1.37{{M_\odot}}$ and a
high moment of inertia, $2\times10^{45}$\,g\,cm$^2$.}

Given these considerations, we search for \ac{BNS} systems with both masses
$\in [1,3] M_{\odot}$ and component dimensionless spins $< 0.05$. We have found
that \ac{BNS} systems with spins $< 0.4$ are generally still recovered well even
though they are not explicitly covered by our search space. Increasing the
search space to include \ac{BNS} systems with spins $< 0.4$ was found to
not improve overall search sensitivity~\citep{nitzthesis}.

\ac{NSBH} systems are thought to be efficiently formed in one of two ways: either
through the stellar evolution of field binaries or through dynamical 
capture of a \ac{NS} by a \ac{BH}~\citep{Grindlay:2005ym,Sadowski:2007dz,Lee:2009ca,Benacquista:2011kv}. 
Though no \ac{NSBH} systems are known to 
exist, one likely progenitor has been observed, Cyg
X-3~\citep{Belczynski:2012jc}. 

Measurements of galactic stellar mass
\acp{BH} in X-ray binaries yield \ac{BH} masses
$5 \le M_{\rm BH}/{M_{\odot}}\le 24$~\citep{Farr:2010tu,Ozel:2010su,Merloni:2008tj,Wiktorowicz:2013dua}.
Extragalactic high-mass X-ray binaries, such as IC10 X-1 and NGC300 X-1 suggest
BH masses of $20-30\,{M_{\odot}}$. Advanced \ac{LIGO} has observed two definitive \ac{BBH} systems and
constrained the masses of the 4 component \acp{BH} to $36_{-4}^{+5},
29_{-4}^{+4}, 14_{-4}^{+8}$ and $7.5_{-2.3}^{+2.3}\,M_{\odot}$, respectively, and
the masses of the two resulting \acp{BH} to $62_{-4}^{+4}$ and
$21_{-2}^{+6}\,M_{\odot}$. In addition if one assumes that the candidate \ac{BBH} merger LVT151012
was of astrophysical origin than its component \acp{BH} had masses constrained to $23_{-6}^{+16}$
and $13_{-5}^{+4}$ with a resulting \ac{BH} mass of $35_{-4}^{+14}$.
There is an apparent gap of \acp{BH} in the mass range $3$--$5\,
{M_{\odot}}$, which has been ascribed to the supernova explosion
mechanism~\citep{Belczynski:2011bn,Fryer:2011cx}. However, \acp{BH} formed from stellar
evolution may exist with masses down to $2\,{M_{\odot}}$, especially if they are formed
from matter accreted onto neutron stars \citep{O'Shaughnessy:2005qc}.
Population synthesis models typically allow for stellar-mass \ac{BH} up to
$\sim 80\text{--}100\,{M_{\odot}}$~\citep{Fryer:2011cx,Belczynski:2009xy,Dominik:2012kk};
stellar \acp{BH} with mass above $100\,{M_{\odot}}$ are also conceivable however
\citep{Belczynski:2014iua,deMink:2015yea}. 

X-ray observations of accreting \acp{BH} indicate a broad
distribution of \ac{BH} spin~\citep{Miller:2009cw,Shafee:2005ef,
McClintock:2006xd,Liu:2008tk,Gou:2009ks,Davis:2006cm, Li:2004aq,Miller:2014aaa}. 
Some \acp{BH} observed in X-ray binaries have very large dimensionless spins 
(e.g Cygnus X-1 at $>0.95$ \citep{2012MNRAS.424..217F,Gou:2011nq}), while others
could have much lower spins ($\sim 0.1$)~\citep{McClintock:2011zq}. Measured
\ac{BH} spins in high-mass X-ray binary systems tend to have large values ($>0.85$), and
these systems are more likely to be progenitors of \ac{NSBH}
binaries~\citep{McClintock:2013vwa}.  Isolated \ac{BH} spins are only constrained
by the relativistic Kerr bound ${\chi} \leq 1$~\cite{MTW}. 
LIGO's observations of merging binary \ac{BH} systems yield 
weak constraints on component spins \citep{TheLIGOScientific:2016wfe,Abbott:2016nmj,TheLIGOScientific:2016pea}.
The microquasar XTE J1550-564~\citep{Steiner:2011vr} and population synthesis
models~\citep{Fragos:2010tm} indicate small spin-orbit misalignment in
field binaries.
Dynamically formed \ac{NSBH} systems, in contrast, are expected to 
have no correlation between the spins and the orbit. 

We search for \ac{NSBH} systems with NS mass
$\in [1,3] M_{\odot}$, NS dimensionless spins $< 0.05$, BH mass $\in [2,99] M_{\odot}$
and BH spin magnitude $< 0.99$. Current search techniques are restricted to
waveform models where the spins are (anti-)aligned with the orbit~\citep{Messick:2016aqy,Usman:2015kfa},
although methods to extend this to generic spins are being explored~\citep{Harry:2016ijz}.
Nevertheless, aligned-spin searches have been shown to have good sensitivity
to systems with generic spin orientations in \ac{O1}~\citep{Canton:2014uja,Harry:2016ijz}. An
additional search for BBH systems with total mass greater than 100 $M_{\odot}$
is also being performed, the results of which will be reported in a future publication.

\section{Search Description}
\label{sec:search_description}

\begin{figure}[t]
\centering
\includegraphics[width=0.45\textwidth]{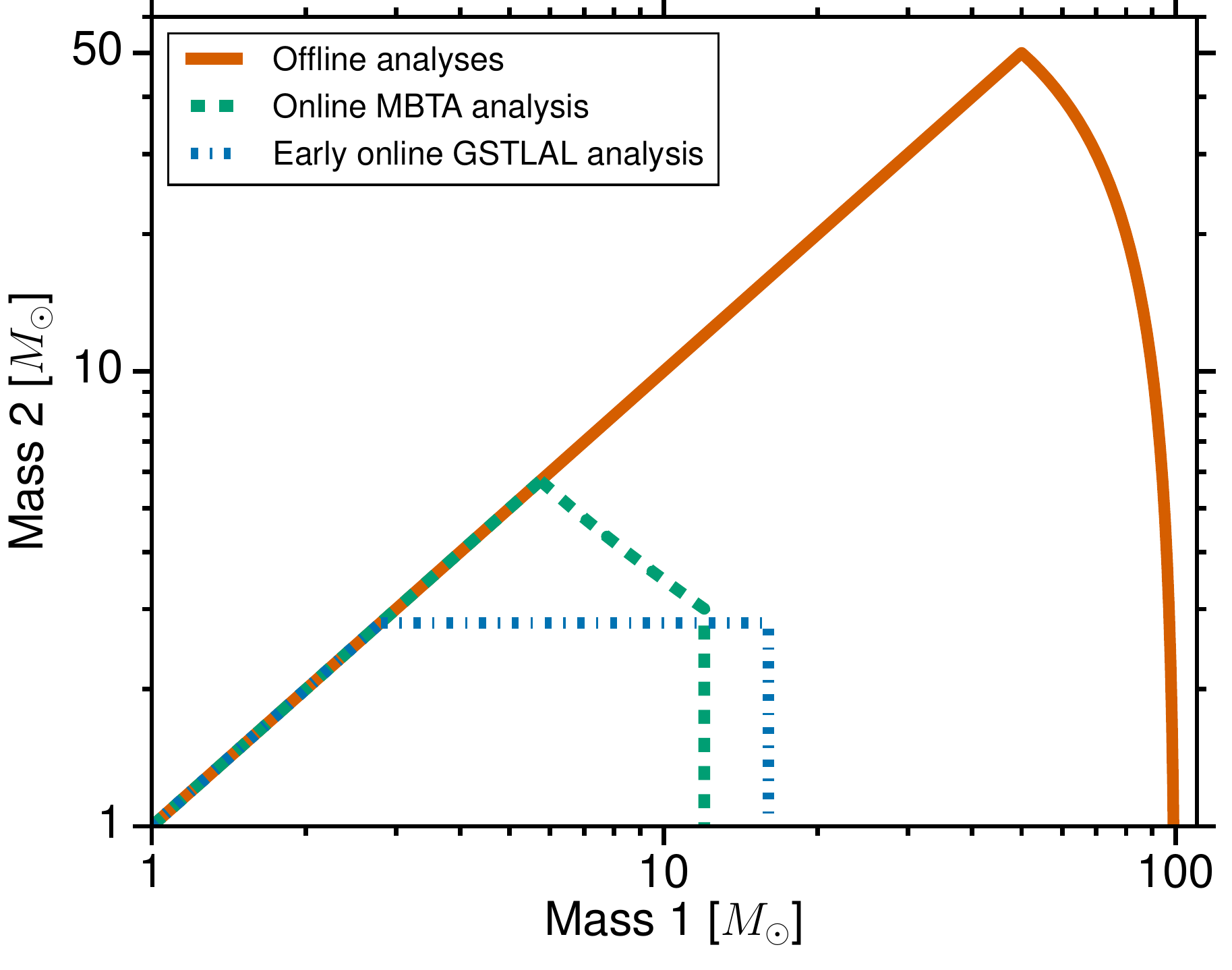}
\caption{\label{fig:banks}The range of template mass parameters considered for the
three different template banks used in the search.
The offline analyses and online \gstlal\ after December 23, 2015, used the largest
bank up to total masses of $100 M_{\odot}$.
The online \mbta\ bank covered primary masses below $12 M_{\odot}$
and chirp masses\textsuperscript{\ref{foot:note1}} below
$5 M_{\odot}$. The early online \gstlal\ bank up to December 23, 2015, covered primary
masses up to $16 M_{\odot}$ and secondary masses up to $2.8 M_{\odot}$.
The spin ranges are not shown here but are discussed in the text. }
\end{figure}

To observe compact binary coalescences in data taken from Advanced \ac{LIGO} we use
matched-filtering against models of compact binary merger \ac{GW} signals~\citep{Wainstein}.
Matched-filtering has long been the primary tool for modeled \ac{GW} searches~\citep{Abbott:2003pj, Colaboration:2011np}.
As the emitted \ac{GW} signal varies significantly over the range of masses and spins
in the \ac{BNS} and \ac{NSBH} parameter space, the matched-filtering process must be repeated
over a large set of filter waveforms, or ``template bank''~\citep{Owen:1998dk}.
The ranges of masses considered in the searches are shown in Figure~\ref{fig:banks}.
The matched-filter process is conducted independently for each of the two \ac{LIGO} observatories before searching
for any potential \ac{GW} signals observed at both observatories with the
same masses and spins and within the expected light travel time delay.
A summary statistic is then assigned to each coincident event based on the estimated
rate of false alarms produced by the search background that would be more significant than the event.

\ac{BNS} and \ac{NSBH} mergers are prime candidates not only for observation with
\ac{GW} facilities, but also for coincident observation with \ac{EM}
observatories~\citep{Eichler:1989ve, Hansen:2000am, Narayan:1992iy, Li:1998bw, Nakar:2007yr, Metzger:2011bv, Nakar:2011cw, Berger:2013jza, Zhang:2013lta, Fong:2015oha}.
We have a long history of working with the Fermi, Swift and IPN \ac{GRB} teams
to perform sub-threshold searches of \ac{GW} data in a narrow window around
the time of observed \acp{GRB}~\citep{Abbott:2005yy, Abbott:2007rh, Abadie:2012bz, Briggs:2012ce}.
Such a search is currently being performed on \ac{O1} data and will be reported in a forthcoming
publication.
In \ac{O1} we also aimed to rapidly alert \ac{EM}
partners if a \ac{GW} observation was made~\citep{Abbott:2016gcq}.
Therefore it was critical for us to run ``online'' searches
to identify potential \ac{BNS} or \ac{NSBH} mergers within a timescale of minutes after the data
is taken, to give \ac{EM} partners the best chance to perform a coincident
observation.

Nevertheless, analyses running with minute latency do not have access to full
data-characterization studies, which can take weeks to perform, or to data with the most complete
knowledge about calibration and associated uncertainties. Additionally, in rare
instances, online analyses may fail to analyse stretches of data due to computational failure. Therefore it is also
important to have an ``offline'' search, which performs the most sensitive search possible
for \ac{BNS} and \ac{NSBH} sources.
We give here a brief description
of both the offline and online searches, referring to other works to give more details
when relevant.

  \subsection{Offline Search}
  \label{ssec:offline_searches}
  \input{offline_search.tex}

  \subsection{Online Search}
  \label{ssec:online_searches}
  \input{online_search.tex}

  \subsection{Dataset}
  \label{ssec:dataset}
  \input{dataset.tex}

\section{Search Results}
\label{sec:search_results}
\input{non_detection.tex}

\section{Rates}
\label{sec:rates}
\input{rates.tex}

\section{Astrophysical Interpretation}
\label{sec:astrophys_interp}

\begin{figure}[t]
\centering
\includegraphics[width=0.5\textwidth]{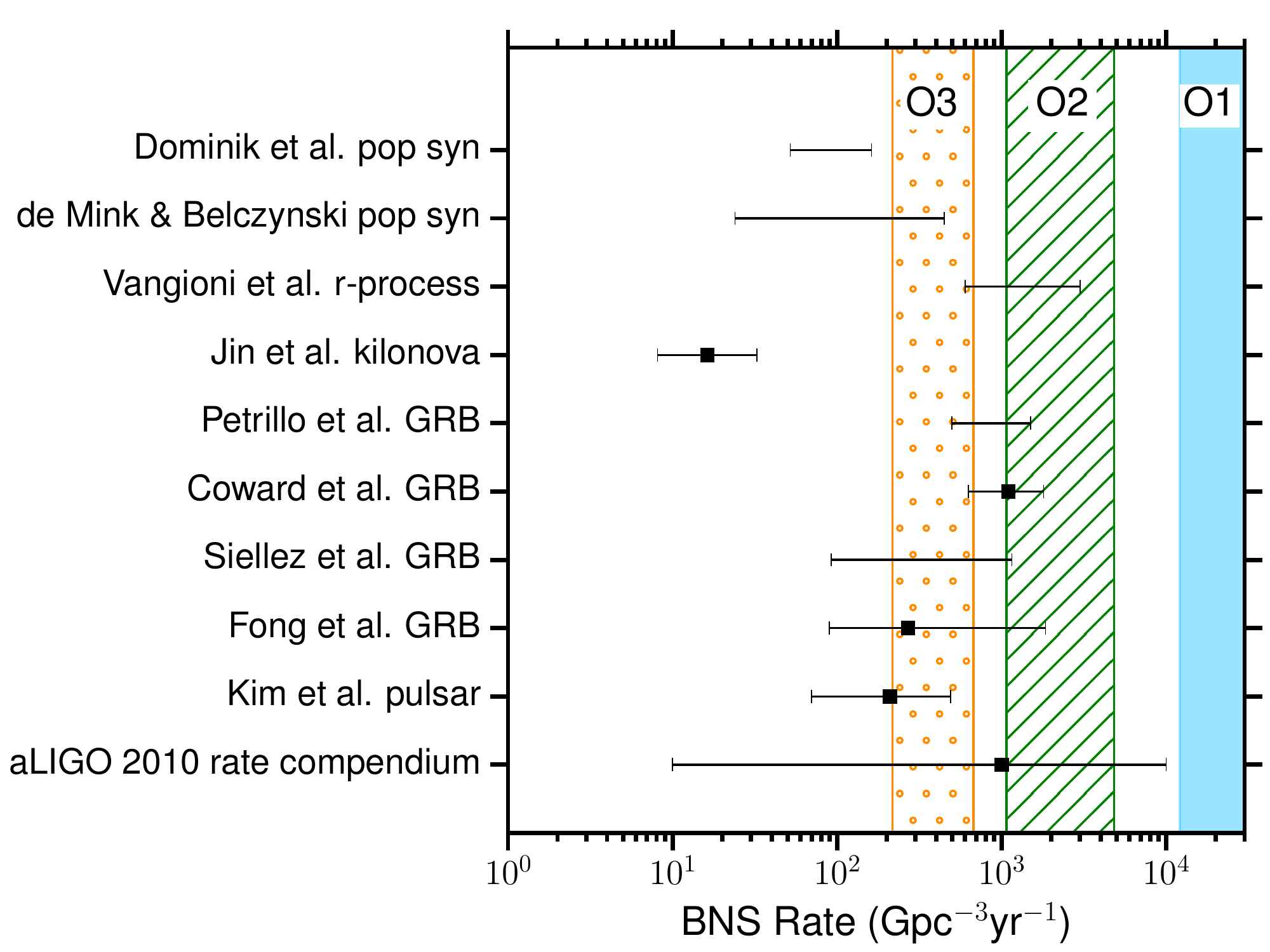}
\caption{\label{fig:ratecomparebns} A comparison of the \ac{O1} 90\% upper limit on the
\ac{BNS} merger rate to other rates discussed in the text \protect\citep{
Abadie:2010cf, Kim:2013tca, Fong:2015oha, Siellez:2013hia, Coward:2012gn,
Petrillo:2012ij, Jin:2015txa, Vangioni:2015ofa, deMink:2015yea, Dominik:2014yma}.  The region excluded by the low-spin \ac{BNS} rate limit is 
shaded in blue.  Continued non-detection in O2 (slash) and O3 (dot) with higher
sensitivities and longer operation time would imply stronger upper limits.  The
O2 and O3 \ac{BNS} ranges are assumed to be 1-1.9 and 1.9-2.7 times larger than
\ac{O1}.  The operation times are assumed to be 6 and 9
months~\citep{Aasi:2013wya} with a duty cycle equal to that of \ac{O1} ($\sim$ 40\%).} 
\end{figure}

\begin{figure}[t]
\centering
\includegraphics[width=0.5\textwidth]{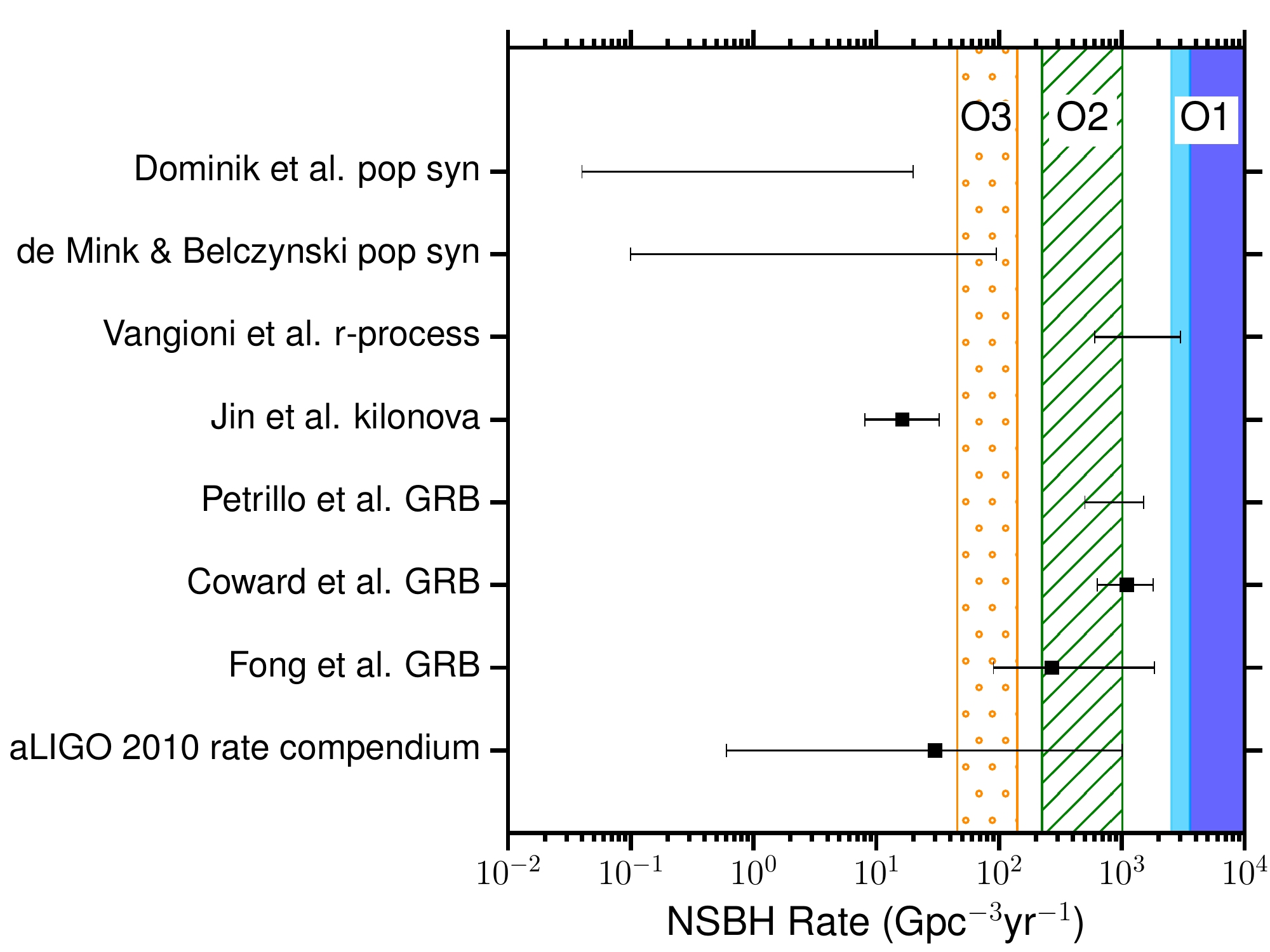}
\caption{\label{fig:ratecomparensbh} A comparison of the \ac{O1} 90\% upper limit on
the \ac{NSBH} merger rate to other rates discussed in
the text \protect\citep{Abadie:2010cf, Fong:2015oha, Coward:2012gn,
Petrillo:2012ij, Jin:2015txa, Vangioni:2015ofa, deMink:2015yea, Dominik:2014yma}.
The dark blue region assumes a \ac{NSBH} population with masses 5--1.4 $M_{\odot}$ and the
light blue region assumes a \ac{NSBH} population with masses 10--1.4 $M_{\odot}$.
Both assume an isotropic spin distribution.
Continued non-detection in O2 (slash) and O3 (dot) with higher sensitivities and longer
operation time would imply stronger upper limits (shown for 10--1.4 $M_{\odot}$ \ac{NSBH}
systems). 
The O2 and O3 ranges are assumed to be 1-1.9 and 1.9-2.7 times larger than
\ac{O1}. 
The operation times are assumed to be 6 and 9 months~\citep{Aasi:2013wya}
with a duty cycle equal to that of \ac{O1} ($\sim$ 40\%).}
\end{figure}

\begin{figure}[t]
\centering
\includegraphics[width=0.5\textwidth]{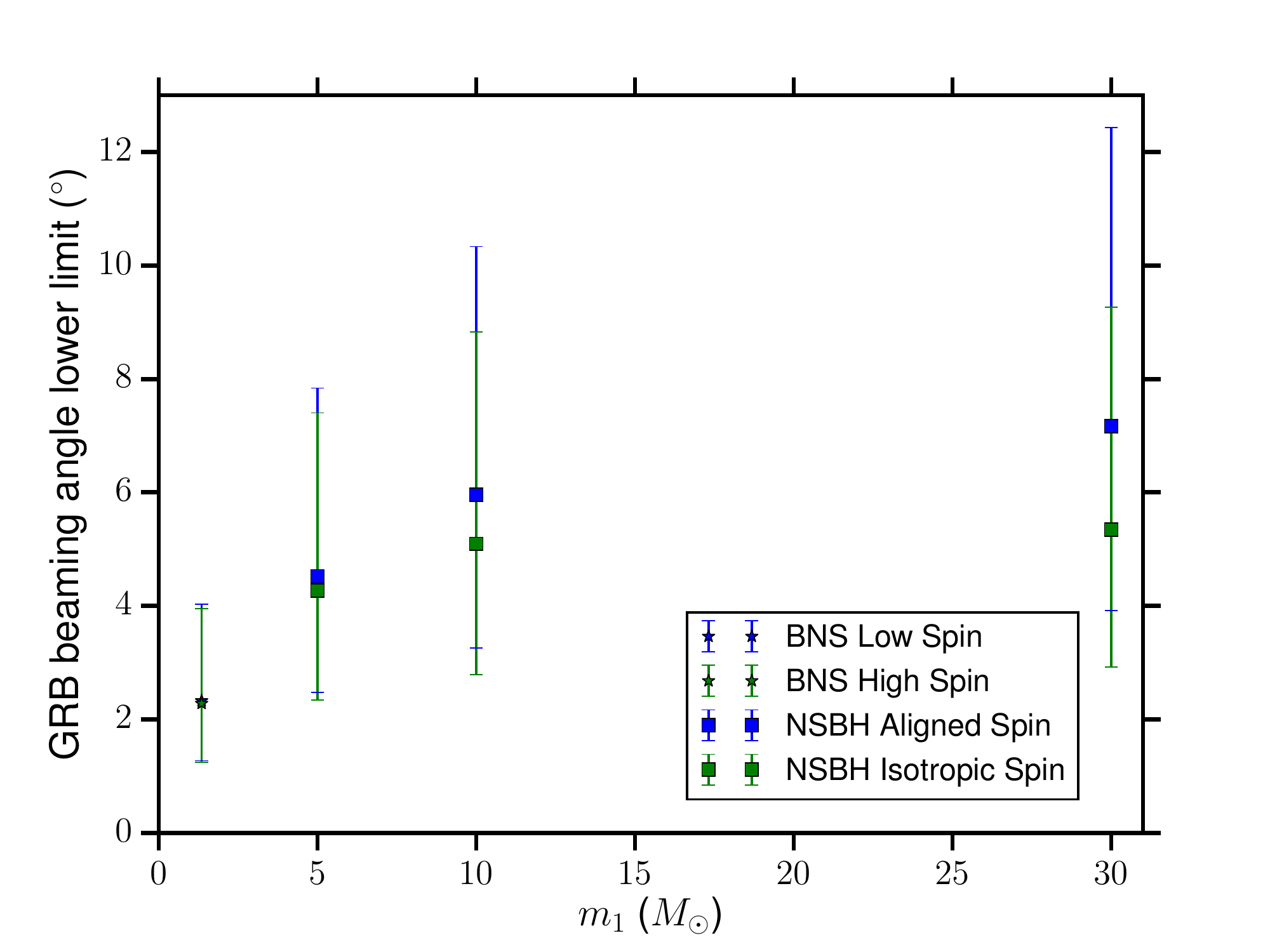}
\caption{\label{fig:beaming} Lower limit on the beaming angle of short
\acp{GRB}, as a function of the mass of the primary BH or
NS, $m_1$. We take the appropriate  90\% rate upper limit from this paper, 
assume all short \acp{GRB} are produced by each case in turn, and assume all
have the same beaming angle $\theta_j$. The limit is calculated using an
observed short \ac{GRB} rate of
$10^{+20}_{-7}$Gpc$^{-3}$ yr$^{-1}$
and the ranges shown on the plot reflect the uncertainty in this observed rate.
For \ac{BNS}, $m_2$ comes from a Gaussian distribution centered on $1.35M_\odot$, and
for \ac{NSBH} it is fixed to $1.4M_\odot$.}
\end{figure}

We can compare our upper limits with rate predictions for compact object mergers
involving \acp{NS}, shown for \ac{BNS} in Figure~\ref{fig:ratecomparebns} and for \ac{NSBH} in
Figure~\ref{fig:ratecomparensbh}. A wide range of predictions derived from population
synthesis and from binary pulsar observations were reviewed in 2010 to produce rate estimates
for canonical 
$1.4\,{{M_{\odot}}}$ \acp{NS} and $10\,{{M_{\odot}}}$ \acp{BH}~\citep{Abadie:2010cf}. We
additionally include some more recent estimates from population synthesis for
both \ac{NSBH} and \ac{BNS} \citep{Dominik:2014yma,Belczynski:2015tba,deMink:2015yea} and
binary pulsar observations for \ac{BNS} \citep{Kim:2013tca}. 

We also compare our upper limits for \ac{NSBH} and \ac{BNS} systems to beaming-corrected
estimates of short \ac{GRB} rates in the local universe. Short \acp{GRB} are 
considered likely to be produced by the merger of compact
binaries that include \acp{NS}, i.e. \ac{BNS} or \ac{NSBH}
systems~\citep{Berger:2013jza}. The rate of short \acp{GRB} can 
predict the rate of progenitor mergers %\ac{BNS} and/or \ac{NSBH} systems
\citep{Coward:2012gn,Petrillo:2012ij,Siellez:2013hia,Fong:2015oha}.
For \ac{NSBH}, systems with small \ac{BH} masses are considered more likely to be able to
produce short \acp{GRB} (e.g.~ \citep{Duez:2009yz,Giacomazzo:2012zt,Pannarale:2015jia}), so we compare to our
$5 M_{\odot}$--$1.4 M_{\odot}$
\ac{NSBH} rate constraint. The observation of a kilonova is also considered to be an
indicator of a binary merger~\citep{Metzger:2011bv}, and an estimated kilonova rate
gives an additional lower bound on compact binary mergers~\citep{Jin:2015txa}.

Finally, some recent work has used the idea that mergers involving \acp{NS}
are the primary astrophysical source of r-process
elements \citep{1974ApJ...192L.145L,Qian:2007vq} to constrain the rate of such
mergers from nucleosynthesis \citep{Bauswein:2014vfa,Vangioni:2015ofa}, and we
include rates from \citep{Vangioni:2015ofa} for comparison.

While limits from \ac{O1} are not yet in tension with astrophysical models, scaling
our results to current expectations for advanced \ac{LIGO}'s next two observing runs,
O2 and O3 \citep{Aasi:2013wya}, suggests that significant constraints or
observations of \ac{BNS} or \ac{NSBH} mergers are possible in the next two years.

Assuming that short \acp{GRB} are produced by \ac{BNS} or \ac{NSBH}, but
without using beaming angle estimates, we can constrain the beaming angle of the jet
of gamma rays emitted from these \acp{GRB} by comparing the rates of
\ac{BNS}/\ac{NSBH} mergers and the rates of
short \acp{GRB}~\citep{Chen:2012qh}. 
For simplicity, we assume here that all short \acp{GRB} are associated with \ac{BNS}
or \ac{NSBH} mergers; the true fraction will
depend on the emission mechanism.  The short \ac{GRB} rate $R_{GRB}$, the merger rate
$R_{merger}$, and the beaming angle $\theta_j$ are then related by 
\begin{linenomath*}
\begin{equation}\label{eq:beaming}
\cos \theta_j = 1 - \frac{R_{\mathrm{GRB}}}{R_{\mathrm{merger}}}
\end{equation}
\end{linenomath*}
We take $R_{GRB}=10^{+20}_{-7}$Gpc$^{-3}$
yr$^{-1}$~\citep{Coward:2012gn,Nakar:2005bs}. 
Figure~\ref{fig:beaming} shows the resulting \ac{GRB} beaming lower limits for the
90\% \ac{BNS} and \ac{NSBH} rate upper limits. 
With our assumption that all short \ac{GRB}s are produced by a single progenitor
class, the constraint is tighter for \ac{NSBH} with larger
\ac{BH} mass.
Observed \ac{GRB} beaming angles are in the range of
$3-25^{\circ}$~\citep{Fox:2005kv,Fong:2015oha,Grupe:2006uc,Soderberg:2006bn,2013ApJ...766...41S,2012ApJ...756...63M,2011A&A...531L...6N}. 
Compared to the lower limit derived from our non-detection, these \ac{GRB}
beaming observations start to confine the fraction of \ac{GRB}s that can be
produced by higher-mass NSBH as progenitor systems.
Future constraints could also come from \ac{GRB} and \ac{BNS} or \ac{NSBH} joint
detections~\citep{Dietz:2010eh,Regimbau:2014nxa, Clark:2014jpa}. 

\section{Conclusion}
\label{sec:conclusion}

We report the non-detection of \ac{BNS} and \ac{NSBH} mergers in advanced \ac{LIGO}'s first observing run.
Given the sensitive volume of Advanced \ac{LIGO} to such systems we are able to place 90\%
confidence upper limits on the rates of \ac{BNS} and \ac{NSBH} mergers, improving upon limits
obtained from Initial \ac{LIGO} and Initial Virgo by roughly an order of magnitude.
Specifically we constrain the merger rate of \ac{BNS} systems with component masses of $1.35\pm0.13M_{\odot}$ 
to be less than \MainBNSULPyCBCHighSpin~Gpc$^{-3}$~yr$^{-1}$. We also constrain
the rate of \ac{NSBH} systems with NS masses of $1.4M_\odot$ and BH masses of at least $5M_{\odot}$ to be less than \MainNSBHULPyCBCFiveAligned~Gpc$^{-3}$~yr$^{-1}$ if
one considers a population where the component spins are (anti-)aligned with
the orbit, and less than \MainNSBHULPyCBCFiveIso~Gpc$^{-3}$~yr$^{-1}$ if one considers an isotropic distribution
of component spin directions.

We compare these upper limits with existing astrophysical rate models and find that the
current upper limits are in conflict with only the most optimistic models of the merger
rate. However, we expect that during the next two observing runs, O2 and O3, we will
either make observations of \ac{BNS} and \ac{NSBH} mergers or start placing significant constraints
on current astrophysical rates. Finally, we have explored the implications of this non-detection on
the beaming angle of short \acp{GRB}. We find that, if one assumes that all \acp{GRB}
are produced by \ac{BNS} mergers, then the opening angle of gamma-ray radiation must be larger
than \GRBBNSBeamingAngleConstraint; or larger than \GRBNSBHFiveBeamingAngleConstraint\ if
one assumes all \acp{GRB} are produced by \ac{NSBH} mergers.

%% ________________________________________________________
\section*{acknowledgments}

The authors gratefully acknowledge the support of the United States
National Science Foundation (NSF) for the construction and operation of the
LIGO Laboratory and Advanced LIGO as well as the Science and Technology Facilities Council (STFC) of the
United Kingdom, the Max-Planck-Society (MPS), and the State of
Niedersachsen/Germany for support of the construction of Advanced LIGO 
and construction and operation of the GEO600 detector. 
Additional support for Advanced LIGO was provided by the Australian Research Council.
The authors gratefully acknowledge the Italian Istituto Nazionale di Fisica Nucleare (INFN),  
the French Centre National de la Recherche Scientifique (CNRS) and
the Foundation for Fundamental Research on Matter supported by the Netherlands Organisation for Scientific Research, 
for the construction and operation of the Virgo detector
and the creation and support  of the EGO consortium. 
The authors also gratefully acknowledge research support from these agencies as well as by 
the Council of Scientific and Industrial Research of India, 
Department of Science and Technology, India,
Science \& Engineering Research Board (SERB), India,
Ministry of Human Resource Development, India,
the Spanish Ministerio de Econom\'ia y Competitividad,
the Conselleria d'Economia i Competitivitat and Conselleria d'Educaci\'o, Cultura i Universitats of the Govern de les Illes Balears,
the National Science Centre of Poland,
the European Commission,
the Royal Society, 
the Scottish Funding Council, 
the Scottish Universities Physics Alliance, 
the Hungarian Scientific Research Fund (OTKA),
the Lyon Institute of Origins (LIO),
the National Research Foundation of Korea,
Industry Canada and the Province of Ontario through the Ministry of Economic Development and Innovation, 
the Natural Science and Engineering Research Council Canada,
Canadian Institute for Advanced Research,
the Brazilian Ministry of Science, Technology, and Innovation,
Funda\c{c}\~ao de Amparo \`a Pesquisa do Estado de S\~ao Paulo (FAPESP),
Russian Foundation for Basic Research,
the Leverhulme Trust, 
the Research Corporation, 
Ministry of Science and Technology (MOST), Taiwan
and
the Kavli Foundation.
The authors gratefully acknowledge the support of the NSF, STFC, MPS, INFN, CNRS and the
State of Niedersachsen/Germany for provision of computational resources.

%% ________________________________________________________

\bibliography{references}

%% +++++++++++++++++++++++++++++++++++++++++++++++++++++++
\end{document}

%% file: macros.tex
\newcommand{\macro}[1]{{#1}} 

\newcommand\OoneSTART{\macro{September 12, 2015}}
\newcommand\OoneEND{\macro{January 19, 2016}}

\newcommand\OoneOfflineAnalysableHTimeSeconds{\macro{76.7~days}}
\newcommand\OoneOfflineAnalysableLTimeSeconds{\macro{65.8~days}}

\newcommand\OoneOfflineAnalysableTimeSeconds{\macro{49.0~days}}

\newcommand\OoneOfflineAnalysableCatTwoDiffSeconds{\macro{0.5~days}}

\newcommand\OoneOnlineAnalysableTimeSeconds{\macro{52.2~days}}

\newcommand\OoneOnlineAnalysedMBTATimeSeconds{\macro{50.5~days}}
\newcommand\OoneOnlineAnalysedMBTATimePercent{\macro{96.6~\%}}

\newcommand\OoneOnlineAnalysedGSTLALTimeSeconds{\macro{49.4~days}}

\newcommand\OoneOnlineAnalysedGSTLALTimePercent{\macro{94.6~\%}}
\newcommand\OoneOnlineAnalysedBOTHTimeSeconds{\macro{52.0~days}}
\newcommand\OoneOnlineAnalysedBOTHTimePercent{\macro{99.5~\%}}

\newcommand\OoneOnlineTotalMBTAGraceDBEvents{\macro{486}}
\newcommand\OoneOnlineTotalGSTLALGraceDBEvents{\macro{868}}
\newcommand\OoneOnlineTotalEMFollowUpEvents{\macro{8}}

\newcommand\LVBLAHsignificance{\macro{$1.7 \sigma$}}

\newcommand\MainBNSULLowSpin{\macro{\MainBNSULPyCBCLowSpin}}
\newcommand\MainBNSULHighSpin{\macro{\MainBNSULPyCBCHighSpin}}
\newcommand\MainBNSULPyCBCLowSpin{\macro{12,100}}
\newcommand\MainBNSULPyCBCHighSpin{\macro{12,600}}
\newcommand\MainBNSULGstlalLowSpin{\macro{11,500}}
\newcommand\MainBNSULGstlalHighSpin{\macro{12,200}}

\newcommand\SSixULNoSpin{\macro{130,000 $\mathrm{Gpc}^{-3}$ $\mathrm{yr}^{-1}$}}

\newcommand\MainBNSVTPyCBCLowSpin{\macro{$2.09\times 10^{-4}$}}
\newcommand\MainBNSVTPyCBCHighSpin{\macro{$2.00\times 10^{-4}$}}
\newcommand\MainBNSVTGstlalLowSpin{\macro{$2.20\times 10^{-4}$}}
\newcommand\MainBNSVTGstlalHighSpin{\macro{$2.07\times 10^{-4}$}}

\newcommand\MainBNSRangePyCBCLowSpin{\macro{73.2}}
\newcommand\MainBNSRangePyCBCHighSpin{\macro{72.1}}
\newcommand\MainBNSRangeGstlalLowSpin{\macro{73.4}}
\newcommand\MainBNSRangeGstlalHighSpin{\macro{72.0}}

\newcommand\MainNSBHULPyCBCFiveIso{\macro{3,600}}
\newcommand\MainNSBHULPyCBCTenIso{\macro{2,530}}
\newcommand\MainNSBHULPyCBCThirtyIso{\macro{2,300}}
\newcommand\MainNSBHULPyCBCFiveAligned{\macro{3,210}}
\newcommand\MainNSBHULPyCBCTenAligned{\macro{1,850}}
\newcommand\MainNSBHULPyCBCThirtyAligned{\macro{1,280}}
\newcommand\MainNSBHVTPyCBCFiveIso{\macro{$7.01\times 10^{-4}$}}
\newcommand\MainNSBHVTPyCBCTenIso{\macro{$1.00\times 10^{-3}$}}
\newcommand\MainNSBHVTPyCBCThirtyIso{\macro{$1.10\times 10^{-3}$}}
\newcommand\MainNSBHVTPyCBCFiveAligned{\macro{$7.87\times 10^{-4}$}}
\newcommand\MainNSBHVTPyCBCTenAligned{\macro{$1.36\times 10^{-3}$}}
\newcommand\MainNSBHVTPyCBCThirtyAligned{\macro{$1.98\times 10^{-3}$}}
\newcommand\MainNSBHRangePyCBCFiveIso{\macro{110}}
\newcommand\MainNSBHRangePyCBCTenIso{\macro{123}}
\newcommand\MainNSBHRangePyCBCThirtyIso{\macro{127}}
\newcommand\MainNSBHRangePyCBCFiveAligned{\macro{114}}
\newcommand\MainNSBHRangePyCBCTenAligned{\macro{137}}
\newcommand\MainNSBHRangePyCBCThirtyAligned{\macro{155}}

\newcommand\MainNSBHULGstlalFiveIso{\macro{3,270}}
\newcommand\MainNSBHULGstlalTenIso{\macro{2,490}}
\newcommand\MainNSBHULGstlalThirtyIso{\macro{2,800}}
\newcommand\MainNSBHULGstlalFiveAligned{\macro{2,820}}
\newcommand\MainNSBHULGstlalTenAligned{\macro{1,660}}
\newcommand\MainNSBHULGstlalThirtyAligned{\macro{1,270}}
\newcommand\MainNSBHVTGstlalFiveIso{\macro{$7.71\times 10^{-4}$}}
\newcommand\MainNSBHVTGstlalTenIso{\macro{$1.01\times 10^{-3}$}}
\newcommand\MainNSBHVTGstlalThirtyIso{\macro{$9.02\times 10^{-4}$}}
\newcommand\MainNSBHVTGstlalFiveAligned{\macro{$8.96\times 10^{-4}$}}
\newcommand\MainNSBHVTGstlalTenAligned{\macro{$1.52\times 10^{-3}$}}
\newcommand\MainNSBHVTGstlalThirtyAligned{\macro{$1.99\times 10^{-3}$}}
\newcommand\MainNSBHRangeGstlalFiveIso{\macro{112}}
\newcommand\MainNSBHRangeGstlalTenIso{\macro{122}}
\newcommand\MainNSBHRangeGstlalThirtyIso{\macro{118}}
\newcommand\MainNSBHRangeGstlalFiveAligned{\macro{117}}
\newcommand\MainNSBHRangeGstlalTenAligned{\macro{140}}
\newcommand\MainNSBHRangeGstlalThirtyAligned{\macro{153}}

\newcommand\SSixNSBHULFiveSpin{\macro{36,000 $\mathrm{Gpc}^{-3}$ $\mathrm{yr}^{-1}$}}

\newcommand\MainBNSRange{\macro{$\sim$ 70 $\mathrm{Mpc}$}}
\newcommand\MainNSBHRangeFive{\macro{$\sim$ 110 $\mathrm{Mpc}$}}

\newcommand\GRBBNSBeamingAngleConstraint{\macro{${2.3^{+1.7}_{-1.1}}^{\circ}$}}
\newcommand\GRBNSBHFiveBeamingAngleConstraint{\macro{${4.3^{+3.1}_{-1.9}}^{\circ}$}}

%% file: LSC_Feb2016_Virgo_Feb2016-aas.tex
\author{%
B.~P.~Abbott,\altaffilmark{1}  %benjamin.abbott
R.~Abbott,\altaffilmark{1}  %rich.abbott
T.~D.~Abbott,\altaffilmark{2}  %thomas.abbott
M.~R.~Abernathy,\altaffilmark{3}  %matthew.abernathy
F.~Acernese,\altaffilmark{4,5} %fausto.acernese
K.~Ackley,\altaffilmark{6}  %kendall.ackley
C.~Adams,\altaffilmark{7}  %carl.adams
T.~Adams,\altaffilmark{8} %thomas.adams
P.~Addesso,\altaffilmark{9}  %paolo.addesso
R.~X.~Adhikari,\altaffilmark{1}
% ...
}  %john.zweizig

\medskip
\affiliation {${}^{*}$Deceased, March 2016. ${}^{\dag}$Deceased, May 2015. ${}^{\ddag}$Deceased, March 2015.
\\
{(LIGO Scientific Collaboration and Virgo Collaboration)}%
}% 
\medskip

\altaffiltext {1}{LIGO, California Institute of Technology, Pasadena, CA 91125, USA }
\altaffiltext {2}{Louisiana State University, Baton Rouge, LA 70803, USA }
\altaffiltext {3}{American University, Washington, D.C. 20016, USA }
\altaffiltext {4}{Universit\`a di Salerno, Fisciano, I-84084 Salerno, Italy }
\altaffiltext {5}{INFN, Sezione di Napoli, Complesso Universitario di Monte S.Angelo, I-80126 Napoli, Italy }
\altaffiltext {6}{University of Florida, Gainesville, FL 32611, USA }
\altaffiltext {7}{LIGO Livingston Observatory, Livingston, LA 70754, USA }
\altaffiltext {8}{Laboratoire d'Annecy-le-Vieux de Physique des Particules (LAPP), Universit\'e Savoie Mont Blanc, CNRS/IN2P3, F-74941 Annecy-le-Vieux, France }
\altaffiltext {9}{University of Sannio at Benevento, I-82100 Benevento, Italy and INFN, Sezione di Napoli, I-80100 Napoli, Italy }
\altaffiltext {10}{Albert-Einstein-Institut, Max-Planck-Institut f\"ur Gravi\-ta\-tions\-physik, D-30167 Hannover, Germany }
\altaffiltext {11}{Nikhef, Science Park, 1098 XG Amsterdam, The Netherlands }
\altaffiltext {12}{LIGO, Massachusetts Institute of Technology, Cambridge, MA 02139, USA }
\altaffiltext {13}{Instituto Nacional de Pesquisas Espaciais, 12227-010 S\~{a}o Jos\'{e} dos Campos, S\~{a}o Paulo, Brazil }
\altaffiltext {14}{INFN, Gran Sasso Science Institute, I-67100 L'Aquila, Italy }
\altaffiltext {15}{INFN, Sezione di Roma Tor Vergata, I-00133 Roma, Italy }
\altaffiltext {16}{Inter-University Centre for Astronomy and Astrophysics, Pune 411007, India }
\altaffiltext {17}{International Centre for Theoretical Sciences, Tata Institute of Fundamental Research, Bangalore 560012, India }
\altaffiltext {18}{University of Wisconsin-Milwaukee, Milwaukee, WI 53201, USA }
\altaffiltext {19}{Leibniz Universit\"at Hannover, D-30167 Hannover, Germany }
\altaffiltext {20}{Universit\`a di Pisa, I-56127 Pisa, Italy }
\altaffiltext {21}{INFN, Sezione di Pisa, I-56127 Pisa, Italy }
\altaffiltext {22}{Australian National University, Canberra, Australian Capital Territory 0200, Australia }
\altaffiltext {23}{The University of Mississippi, University, MS 38677, USA }
\altaffiltext {24}{California State University Fullerton, Fullerton, CA 92831, USA }
\altaffiltext {25}{LAL, Univ. Paris-Sud, CNRS/IN2P3, Universit\'e Paris-Saclay, Orsay, France }
\altaffiltext {26}{Chennai Mathematical Institute, Chennai 603103, India }
\altaffiltext {27}{Universit\`a di Roma Tor Vergata, I-00133 Roma, Italy }
\altaffiltext {28}{University of Southampton, Southampton SO17 1BJ, United Kingdom }
\altaffiltext {29}{Universit\"at Hamburg, D-22761 Hamburg, Germany }
\altaffiltext {30}{INFN, Sezione di Roma, I-00185 Roma, Italy }
\altaffiltext {31}{Albert-Einstein-Institut, Max-Planck-Institut f\"ur Gravitations\-physik, D-14476 Potsdam-Golm, Germany }
\altaffiltext {32}{APC, AstroParticule et Cosmologie, Universit\'e Paris Diderot, CNRS/IN2P3, CEA/Irfu, Observatoire de Paris, Sorbonne Paris Cit\'e, F-75205 Paris Cedex 13, France }
\altaffiltext {33}{Montana State University, Bozeman, MT 59717, USA }
\altaffiltext {34}{Universit\`a di Perugia, I-06123 Perugia, Italy }
\altaffiltext {35}{INFN, Sezione di Perugia, I-06123 Perugia, Italy }
\altaffiltext {36}{European Gravitational Observatory (EGO), I-56021 Cascina, Pisa, Italy }
\altaffiltext {37}{Syracuse University, Syracuse, NY 13244, USA }
\altaffiltext {38}{SUPA, University of Glasgow, Glasgow G12 8QQ, United Kingdom }
\altaffiltext {39}{LIGO Hanford Observatory, Richland, WA 99352, USA }
\altaffiltext {40}{Wigner RCP, RMKI, H-1121 Budapest, Konkoly Thege Mikl\'os \'ut 29-33, Hungary }
\altaffiltext {41}{Columbia University, New York, NY 10027, USA }
\altaffiltext {42}{Stanford University, Stanford, CA 94305, USA }
\altaffiltext {43}{Universit\`a di Padova, Dipartimento di Fisica e Astronomia, I-35131 Padova, Italy }
\altaffiltext {44}{INFN, Sezione di Padova, I-35131 Padova, Italy }
\altaffiltext {45}{CAMK-PAN, 00-716 Warsaw, Poland }
\altaffiltext {46}{University of Birmingham, Birmingham B15 2TT, United Kingdom }
\altaffiltext {47}{Universit\`a degli Studi di Genova, I-16146 Genova, Italy }
\altaffiltext {48}{INFN, Sezione di Genova, I-16146 Genova, Italy }
\altaffiltext {49}{RRCAT, Indore MP 452013, India }
\altaffiltext {50}{Faculty of Physics, Lomonosov Moscow State University, Moscow 119991, Russia }
\altaffiltext {51}{SUPA, University of the West of Scotland, Paisley PA1 2BE, United Kingdom }
\altaffiltext {52}{University of Western Australia, Crawley, Western Australia 6009, Australia }
\altaffiltext {53}{Department of Astrophysics/IMAPP, Radboud University Nijmegen, P.O. Box 9010, 6500 GL Nijmegen, The Netherlands }
\altaffiltext {54}{Artemis, Universit\'e C\^ote d'Azur, CNRS, Observatoire C\^ote d'Azur, CS 34229, Nice cedex 4, France }
\altaffiltext {55}{Institut de Physique de Rennes, CNRS, Universit\'e de Rennes 1, F-35042 Rennes, France }
\altaffiltext {56}{Washington State University, Pullman, WA 99164, USA }
\altaffiltext {57}{Universit\`a degli Studi di Urbino ``Carlo Bo,'' I-61029 Urbino, Italy }
\altaffiltext {58}{INFN, Sezione di Firenze, I-50019 Sesto Fiorentino, Firenze, Italy }
\altaffiltext {59}{University of Oregon, Eugene, OR 97403, USA }
\altaffiltext {60}{Laboratoire Kastler Brossel, UPMC-Sorbonne Universit\'es, CNRS, ENS-PSL Research University, Coll\`ege de France, F-75005 Paris, France }
\altaffiltext {61}{Carleton College, Northfield, MN 55057, USA }
\altaffiltext {62}{Astronomical Observatory Warsaw University, 00-478 Warsaw, Poland }
\altaffiltext {63}{VU University Amsterdam, 1081 HV Amsterdam, The Netherlands }
\altaffiltext {64}{University of Maryland, College Park, MD 20742, USA }
\altaffiltext {65}{Center for Relativistic Astrophysics and School of Physics, Georgia Institute of Technology, Atlanta, GA 30332, USA }
\altaffiltext {66}{Laboratoire des Mat\'eriaux Avanc\'es (LMA), CNRS/IN2P3, F-69622 Villeurbanne, France }
\altaffiltext {67}{Universit\'e Claude Bernard Lyon 1, F-69622 Villeurbanne, France }
\altaffiltext {68}{Universit\`a di Napoli ``Federico II,'' Complesso Universitario di Monte S.Angelo, I-80126 Napoli, Italy }
\altaffiltext {69}{NASA/Goddard Space Flight Center, Greenbelt, MD 20771, USA }
\altaffiltext {70}{RESCEU, University of Tokyo, Tokyo, 113-0033, Japan. }
\altaffiltext {71}{Tsinghua University, Beijing 100084, China }
\altaffiltext {72}{Texas Tech University, Lubbock, TX 79409, USA }
\altaffiltext {73}{The Pennsylvania State University, University Park, PA 16802, USA }
\altaffiltext {74}{National Tsing Hua University, Hsinchu City, 30013 Taiwan, Republic of China }
\altaffiltext {75}{Charles Sturt University, Wagga Wagga, New South Wales 2678, Australia }
\altaffiltext {76}{West Virginia University, Morgantown, WV 26506, USA }
\altaffiltext {77}{University of Chicago, Chicago, IL 60637, USA }
\altaffiltext {78}{Caltech CaRT, Pasadena, CA 91125, USA }
\altaffiltext {79}{Korea Institute of Science and Technology Information, Daejeon 305-806, Korea }
\altaffiltext {80}{Universit\`a di Roma ``La Sapienza,'' I-00185 Roma, Italy }
\altaffiltext {81}{University of Brussels, Brussels 1050, Belgium }
\altaffiltext {82}{Sonoma State University, Rohnert Park, CA 94928, USA }
\altaffiltext {83}{Center for Interdisciplinary Exploration \& Research in Astrophysics (CIERA), Northwestern University, Evanston, IL 60208, USA }
\altaffiltext {84}{University of Minnesota, Minneapolis, MN 55455, USA }
\altaffiltext {85}{The University of Melbourne, Parkville, Victoria 3010, Australia }
\altaffiltext {86}{Institute for Plasma Research, Bhat, Gandhinagar 382428, India }
\altaffiltext {87}{The University of Sheffield, Sheffield S10 2TN, United Kingdom }
\altaffiltext {88}{The University of Texas Rio Grande Valley, Brownsville, TX 78520, USA }
\altaffiltext {89}{Universit\`a di Trento, Dipartimento di Fisica, I-38123 Povo, Trento, Italy }
\altaffiltext {90}{INFN, Trento Institute for Fundamental Physics and Applications, I-38123 Povo, Trento, Italy }
\altaffiltext {91}{Cardiff University, Cardiff CF24 3AA, United Kingdom }
\altaffiltext {92}{Montclair State University, Montclair, NJ 07043, USA }
\altaffiltext {93}{MTA E\"otv\"os University, ``Lendulet'' Astrophysics Research Group, Budapest 1117, Hungary }
\altaffiltext {94}{National Astronomical Observatory of Japan, 2-21-1 Osawa, Mitaka, Tokyo 181-8588, Japan }
\altaffiltext {95}{School of Mathematics, University of Edinburgh, Edinburgh EH9 3FD, United Kingdom }
\altaffiltext {96}{Indian Institute of Technology, Gandhinagar Ahmedabad Gujarat 382424, India }
\altaffiltext {97}{University of Szeged, D\'om t\'er 9, Szeged 6720, Hungary }
\altaffiltext {98}{Embry-Riddle Aeronautical University, Prescott, AZ 86301, USA }
\altaffiltext {99}{Tata Institute of Fundamental Research, Mumbai 400005, India }
\altaffiltext {100}{INAF, Osservatorio Astronomico di Capodimonte, I-80131, Napoli, Italy }
\altaffiltext {101}{University of Michigan, Ann Arbor, MI 48109, USA }
\altaffiltext {102}{Rochester Institute of Technology, Rochester, NY 14623, USA }
\altaffiltext {103}{NCSA, University of Illinois at Urbana-Champaign, Urbana, Illinois 61801, USA }
\altaffiltext {104}{Universitat de les Illes Balears, IAC3---IEEC, E-07122 Palma de Mallorca, Spain }
\altaffiltext {105}{University of Bia{\l }ystok, 15-424 Bia{\l }ystok, Poland }
\altaffiltext {106}{SUPA, University of Strathclyde, Glasgow G1 1XQ, United Kingdom }
\altaffiltext {107}{IISER-TVM, CET Campus, Trivandrum Kerala 695016, India }
\altaffiltext {108}{Canadian Institute for Theoretical Astrophysics, University of Toronto, Toronto, Ontario M5S 3H8, Canada }
\altaffiltext {109}{Institute of Applied Physics, Nizhny Novgorod, 603950, Russia }
\altaffiltext {110}{Pusan National University, Busan 609-735, Korea }
\altaffiltext {111}{Hanyang University, Seoul 133-791, Korea }
\altaffiltext {112}{University of Adelaide, Adelaide, South Australia 5005, Australia }
\altaffiltext {113}{NCBJ, 05-400 \'Swierk-Otwock, Poland }
\altaffiltext {114}{IM-PAN, 00-956 Warsaw, Poland }
\altaffiltext {115}{Monash University, Victoria 3800, Australia }
\altaffiltext {116}{Seoul National University, Seoul 151-742, Korea }
\altaffiltext {117}{The Chinese University of Hong Kong, Shatin, NT, Hong Kong }
\altaffiltext {118}{University of Alabama in Huntsville, Huntsville, AL 35899, USA }
\altaffiltext {119}{University of Massachusetts-Amherst, Amherst, MA 01003, USA }
\altaffiltext {120}{ESPCI, CNRS, F-75005 Paris, France }
\altaffiltext {121}{Universit\`a di Camerino, Dipartimento di Fisica, I-62032 Camerino, Italy }
\altaffiltext {122}{Southern University and A\&M College, Baton Rouge, LA 70813, USA }
\altaffiltext {123}{College of William and Mary, Williamsburg, VA 23187, USA }
\altaffiltext {124}{Instituto de F\'\i sica Te\'orica, University Estadual Paulista/ICTP South American Institute for Fundamental Research, S\~ao Paulo SP 01140-070, Brazil }
\altaffiltext {125}{University of Cambridge, Cambridge CB2 1TN, United Kingdom }
\altaffiltext {126}{IISER-Kolkata, Mohanpur, West Bengal 741252, India }
\altaffiltext {127}{Rutherford Appleton Laboratory, HSIC, Chilton, Didcot, Oxon OX11 0QX, United Kingdom }
\altaffiltext {128}{Whitman College, 345 Boyer Avenue, Walla Walla, WA 99362 USA }
\altaffiltext {129}{National Institute for Mathematical Sciences, Daejeon 305-390, Korea }
\altaffiltext {130}{Universit\'e de Lyon, F-69361 Lyon, France }
\altaffiltext {131}{Hobart and William Smith Colleges, Geneva, NY 14456, USA }
\altaffiltext {132}{Janusz Gil Institute of Astronomy, University of Zielona G\'ora, 65-265 Zielona G\'ora, Poland }
\altaffiltext {133}{King's College London, University of London, London WC2R 2LS, United Kingdom }
\altaffiltext {134}{Andrews University, Berrien Springs, MI 49104, USA }
\altaffiltext {135}{Universit\`a di Siena, I-53100 Siena, Italy }
\altaffiltext {136}{Trinity University, San Antonio, TX 78212, USA }
\altaffiltext {137}{University of Washington, Seattle, WA 98195, USA }
\altaffiltext {138}{Kenyon College, Gambier, OH 43022, USA }
\altaffiltext {139}{Abilene Christian University, Abilene, TX 79699, USA }

%% file: acronyms.tex
\acrodef{aLIGO}[aLIGO]{Advanced Laser Interferometer Gravitational Wave Observatory}
\acrodef{BBH}[BBH]{binary black-hole}
\acrodef{BH}[BH]{black-hole}
\acrodef{BNS}[BNS]{binary neutron-star}
\acrodef{CBC}[CBC]{compact binary coalescence}
\acrodef{EM}[EM]{electromagnetic}
\acrodef{GraCEDb}[GraCEDb]{gravitational-wave candidate event database}
\acrodef{GRB}[GRB]{gamma-ray burst}
\acrodef{GW}[GW]{gravitational wave}
\acrodef{LIGO}[LIGO]{Laser Interferometer Gravitational Wave Observatory}
\acrodef{NS}[NS]{neutron-star}
\acrodef{NSBH}{neutron-star--black-hole}
\acrodef{O1}[O1]{first observing period}

%% file: offline_search.tex
The offline \ac{CBC} search of the \ac{O1} data set consists of two independently-implemented matched-filter
analyses: \gstlal~\citep{Messick:2016aqy} and
\pycbc~\citep{Usman:2015kfa}.
For detailed descriptions of these analyses and associated methods we refer the reader
to~\citep{Babak:2012zx,Canton:2014ena,Usman:2015kfa} for \pycbc\
and~\citep{Cannon:2011vi, Cannon:2012zt, Privitera:2013xza, Messick:2016aqy} for \gstlal.
We also refer the reader to~\citep{TheLIGOScientific:2016pea, TheLIGOScientific:2016qqj} for a detailed
description of the offline search of the \ac{O1} dataset, here we give only a brief overview.

In contrast to the online search, the offline search uses data produced with
smaller calibration errors~\citep{Abbott:2016jsd}, uses complete information about the instrumental
data quality~\citep{TheLIGOScientific:2016zmo} and ensures that all available data is analysed.
The offline search in \ac{O1} forms a single search targeting \ac{BNS}, \ac{NSBH}, and \ac{BBH} systems. The
waveform filters cover systems with individual
component masses ranging from 1 to 99 $M_{\odot}$, total mass
constrained to less than 100 $M_{\odot}$ (see Figure~\ref{fig:banks}), and component dimensionless spins up to $\pm$ 0.05
for components with mass less than 2 $M_{\odot}$ and $\pm$ 0.99 otherwise~\citep{TheLIGOScientific:2016pea,Capano:2016dsf}.
Waveform filters with total mass less than 4 $M_{\odot}$ (chirp mass less than
$1.73 {M}_{\odot}$\footnote{\label{foot:note1}The
``chirp mass'' is the combination of the two component masses that
\ac{LIGO} is most sensitive to, given by $\mathcal{M} = (m_1 m_2)^{3/5} (m_1 + m_2)^{-1/5}$, where
$m_i$ denotes the two component masses})
for \pycbc\ (\gstlal) are modeled with the
inspiral-only, post-Newtonian, frequency-domain approximant
``TaylorF2''~\citep{Arun:2008kb,Bohe:2013cla,Blanchet:2013haa,Bohe:2015ana,Mishra:2016whh}.
At larger masses it becomes important to also include the merger and ringdown components of the waveform.
There a reduced-order model of the effective-one-body waveform calibrated against numerical
relativity is used~\citep{Taracchini:2013rva,Purrer:2015tud}.

%% file: online_search.tex
The online \ac{CBC} search of the \ac{O1} data also consisted of two analyses;
an online version of \gstlal\,\citep{Messick:2016aqy} and \mbta\,\citep{Adams:2015ulm}.
For detailed descriptions of the \mbta\ analysis we refer the reader to~\citep{Beauville:2007kp,Virgo:2011aa,Adams:2015ulm}.
The bank of waveform filters used by \gstlal\ up to December 23, 2015---and by
\mbta\ for the duration of \ac{O1}---targeted systems that contained at least one \ac{NS}.
Such systems are most likely to have an \ac{EM} counterpart, which would
be powered by the material from a disrupted \ac{NS}.
These sets of waveform filters were constructed using methods described
in~\citep{Brown:2012qf,Harry:2013tca,Pannarale:2014rea}. \gstlal\ chose to cover systems with component masses of
$m_1 \in [1,16] M_{\odot}; m_2 \in [1,2.8] M_{\odot}$ and \mbta\ covered $m_1, m_2 \in [1,12] M_{\odot}$ with
a limit on chirp mass $\mathcal{M} < 5\mathrm{M}_{\odot}$ (see Figure~\ref{fig:banks}). In \gstlal\ component spins were
limited to $\chi_{i} < 0.05$ for $m_{i} < 2.8\mathrm{M}_{\odot}$ and $\chi_{i} < 1$ otherwise, for
\mbta\ $\chi_{i} < 0.05$ for $m_{i} < 2\mathrm{M}_{\odot}$ and $\chi_{i} < 1$ otherwise.
\gstlal\ also chose to limit the template bank to include only systems for which it is possible for
a \ac{NS} to have disrupted during the late inspiral using constraints described
in~\citep{Pannarale:2014rea}. For the \mbta\ search the waveform filters were modelled using the
``TaylorT4'' time-domain, post-Newtonian inspiral approximant~\citep{Buonanno:2009zt}.
For \gstlal\
the TaylorF2 frequency-domain, post-Newtonian waveform approximant was
used~\citep{Arun:2008kb,Bohe:2013cla,Blanchet:2013haa,Bohe:2015ana,Mishra:2016whh}.
All waveform models used in this paper are publicly available
in the \texttt{lalsimulation} repository~\citep{LAL}.\footnote{The
  internal \texttt{lalsimulation} names for the waveforms used as filters
  described in this work are ``TaylorF2'' for the frequency-domain post-Newtonian
  approximant, ``SpinTaylorT4'' for the time-domain
  approximant used by \mbta\ and ``SEOBNRv2\_ROM\_DoubleSpin'' for the
  aligned-spin effective one body waveform.
  In addition, for calculation of rate estimates describe in Section \ref{sec:rates},
  the ``SpinTaylorT4'' model is used to simulate BNS signals and
  ``SEOBNRv3'' is used to simulate NSBH signals.}

After December 23, 2015, and triggered by the discovery of GW150914, the \gstlal\
analysis was extended to cover the same search space---using
the same set of waveform filters---as the offline search~\citep{Capano:2016dsf,TheLIGOScientific:2016pea}.

%% file: dataset.tex
Advanced \ac{LIGO}'s first observing run occurred between \OoneSTART\ and \OoneEND\
and consists of data from the two \ac{LIGO} observatories in Hanford, WA and Livingston, LA.
The LIGO detectors were running stably with roughly 40\% coincident operation, and had
been commissioned to roughly a third of the design sensitivity by the time of the start of O1~\citep{Martynov:2016fzi}.
During this observing run the final offline dataset consisted of \OoneOfflineAnalysableHTimeSeconds\
of analyzable data from the Hanford observatory, and \OoneOfflineAnalysableLTimeSeconds\ of data from the
Livingston observatory. We analyze only times during which \emph{both} observatories
took analyzable data, which is \OoneOfflineAnalysableTimeSeconds. Characterization studies of the analysable
data found \OoneOfflineAnalysableCatTwoDiffSeconds\ of coincident data during which time
there was some identified instrumental problem---known to
introduce excess noise---in at least one of the interferometers~\citep{TheLIGOScientific:2016zmo}.
These times are removed before assessing the significance of events
in the remaining analysis time. Some additional time is not analysed because
of restrictions on the minimal length of data segments and because of data lost
at the start and end of those segments~\citep{TheLIGOScientific:2016qqj, TheLIGOScientific:2016pea}.
These requirements are slightly different between the two offline analyses and \pycbc\
analysed 46.1 days of data while \gstlal\ analysed 48.3 days of data.

The data available to the online analyses are not exactly the same as that
available to the offline analyses. Some data were not available online due to (for
example) software
failures, and can later be made available for offline analysis. In contrast,
some data identified as analysable for the online codes may later be identified as invalid as the result
of updated data-characterization studies or because of problems in the calibration of the data.
During \ac{O1} a total of \OoneOnlineAnalysableTimeSeconds\
of coincident data was made available for online analysis.
Of this coincident online data \mbta\ analysed
\OoneOnlineAnalysedMBTATimeSeconds\ (\OoneOnlineAnalysedMBTATimePercent) and
\gstlal\ analysed
\OoneOnlineAnalysedGSTLALTimeSeconds\ (\OoneOnlineAnalysedGSTLALTimePercent).
A total of \OoneOnlineAnalysedBOTHTimeSeconds\ (\OoneOnlineAnalysedBOTHTimePercent)
of data was analysed by at least one of the online analyses.

%% file: non_detection.tex
\begin{figure}[t]
  \centering
  \includegraphics[width=0.45\textwidth]{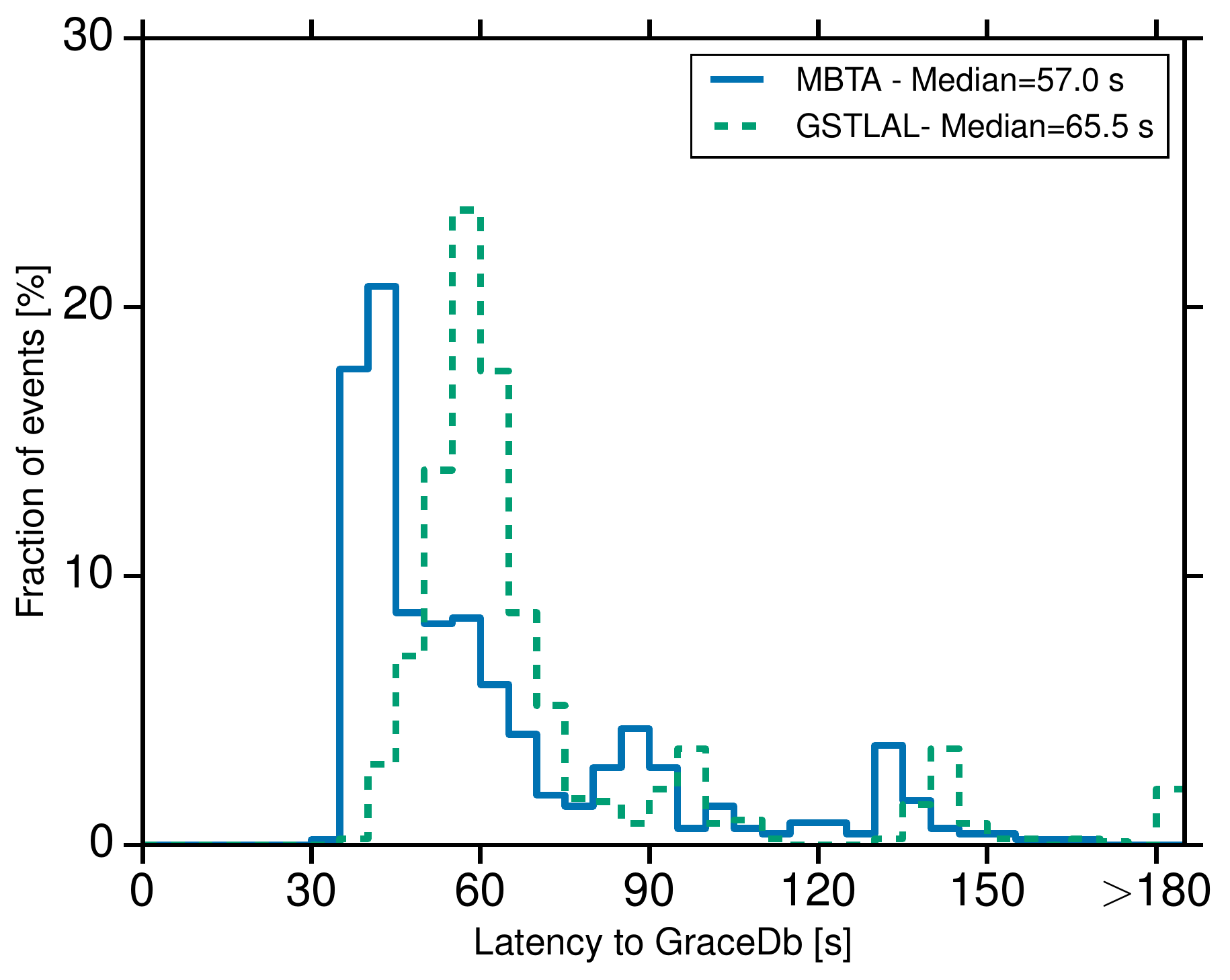}
  \caption{\label{fig:latency} Latency of the online searches during O1.
           The latency is measured as the time between the event arriving at Earth
           and time at which the event is uploaded to \ac{GraCEDb}.}
\end{figure}

The offline search, targeting \ac{BBH} as well as \ac{BNS} and \ac{NSBH} mergers, identified
two signals with $> 5 \sigma$ confidence in the \ac{O1} dataset~\citep{Abbott:2016blz,Abbott:2016nmj}. A third signal was
also identified with \LVBLAHsignificance\ confidence~\citep{TheLIGOScientific:2016pea, TheLIGOScientific:2016qqj}. Subsequent
parameter inference on all three of these events has determined that, to very high
confidence, they were not produced by a \ac{BNS} or \ac{NSBH} merger~\citep{TheLIGOScientific:2016wfe, TheLIGOScientific:2016pea}. No other events
are significant with respect to the noise background in the offline search~\citep{TheLIGOScientific:2016pea}, and
we therefore state that no \ac{BNS} or \ac{NSBH} mergers were observed.

The online search identified a total of \OoneOnlineTotalEMFollowUpEvents\ unique \ac{GW} candidate events
with a false-alarm rate (FAR) less than
$6\, \mathrm{yr}^{-1}$.
Events with a FAR less than this are sent to electromagnetic partners if they
pass event validation.
Six of the events were rejected during the event validation as they were associated
with known non-Gaussian behavior in one of the observatories. Of the remaining
events, one was the \ac{BBH} merger GW151226 reported in~\citep{Abbott:2016nmj}. The second
event identified by \gstlal\ was only narrowly below the FAR threshold, with a FAR of $3.1\, \mathrm{yr}^{-1}$.
This event was also detected by \mbta\ with a higher FAR of $35\, \mathrm{yr}^{-1}$.
This is consistent with noise in the online searches and the candidate event was later
identified to have a false alarm rate of $190\, \mathrm{yr}^{-1}$
in the offline \gstlal\ analysis.
Nevertheless, the event passed all event validation and was released for \ac{EM} follow-up observations,
which showed no significant counterpart. The results
of the \ac{EM} follow-up program are discussed in more detail in~\citep{Abbott:2016gcq}.

All events identified by the \gstlal\ or \mbta\ online analyses with a false alarm rate of less than
$3200 \, \mathrm{yr}^{-1}$
are uploaded to an internal database known
as the \acf{GraCEDb}~\citep{gracedb}.
In total \OoneOnlineTotalMBTAGraceDBEvents\ events were uploaded from \mbta\
and \OoneOnlineTotalGSTLALGraceDBEvents\ from \gstlal.
We can measure the latency of the online pipelines from the time between the
inferred arrival time of each event at the Earth and the time at which the
event is uploaded to \ac{GraCEDb}. This latency is illustrated in Fig.~\ref{fig:latency},
where it can be seen that both online pipelines acheived median latencies on
the order of one minute. We note that \gstlal\ uploaded twice as many events as \mbta\
because of a difference in how the FAR was defined. The FAR reported by \mbta\ was defined
relative to the rate of coincident data such that an event with a FAR of $1 \, \mathrm{yr}^{-1}$
is expected to occur once in a year of coincident data. The FAR reported by \gstlal\ was defined 
relative to wall-clock time such that an event with a FAR of $1 \, \mathrm{yr}^{-1}$
is expected to occur once in a calendar year. In the following section we use the \mbta\
definition of FAR when computing rate upper limits.

%% file: rates.tex
\subsection{Calculating upper limits}
\label{ssec:upper_limits}

Given no evidence for \ac{BNS} or \ac{NSBH} coalescences during \ac{O1}, we seek to place an upper limit
on the astrophysical rate of such events.
The expected number of observed events $\Lambda$ in a given analysis can be
related to the astrophysical rate of coalescences for a given source $R$ by
\begin{linenomath*}
\begin{equation}
\Lambda = R { \langle VT \rangle}.
\end{equation}
\end{linenomath*}
Here, $\langle VT \rangle$ is the space-time volume that the detectors are sensitive
to---averaged over space, observation time, and the parameters of the source population
of interest. The likelihood for finding zero observations in the data $s$ follows the
Poisson distribution for zero events $p(s | \Lambda) = e^{-\Lambda}$. Bayes' theorem then
gives the posterior for $\Lambda$
\begin{linenomath*}
\begin{equation}
p(\Lambda | s) \propto p(\Lambda) e^{-\Lambda},
\label{eq:lambdapost}
\end{equation}
\end{linenomath*}
where $p(\Lambda)$ is the prior on $\Lambda$.

Searches of Initial \ac{LIGO} and Initial Virgo data used a uniform prior on
$\Lambda$~\citep{Colaboration:2011np} but included prior information from previous searches. 
For the \ac{O1} \ac{BBH} search, however, a Jeffreys prior of
$p(\Lambda) \propto 1/\sqrt{\Lambda}$ for the Poisson likelihood was used~\citep{Farr:2013yna,Abbott:2016nhf,TheLIGOScientific:2016pea}.
A Jeffreys prior has the convenient property that the resulting posterior is invariant under a change in parametrization. 
However, for consistency with past BNS and NSBH results we will primarily use a uniform prior, and note that a Jeffreys
prior generally predicts a rate upper limit that is $\sim 40$\% smaller. 
We do not include additional prior information because the sensitive $\langle VT \rangle$ from 
all previous runs is an order of magnitude smaller than that of \ac{O1}.
We estimate $\langle VT \rangle$ by adding a large number of
simulated waveforms sampled from an astrophysical population into the data. 
These simulated signals
are recovered with an estimate of the FAR using the offline analyses.
Monte-Carlo integration methods are then utilized to estimate
the sensitive volume to which the detectors can recover gravitational-wave signals
below a chosen FAR threshold, which in this paper we will choose to be $0.01 \mathrm{yr}^{-1}$.
This threshold is low enough that only signals that are likely to be true events are counted as found, and we 
note that varying this threshold in the range 0.0001--1~yr$^{-1}$ only changes the calculated $\langle VT \rangle$ 
by about $\pm 20\%$.

Calibration uncertainties lead to a difference between the amplitude of simulated
waveforms and the amplitude of real waveforms with the same luminosity distance $d_L$.
During \ac{O1}, the $1\sigma$ uncertainty in the strain amplitude was 6\%, resulting
in an 18\% uncertainty in the measured $\langle VT \rangle$. Results presented here
also assume that injected waveforms are accurate
representations of astrophysical sources. We use a time-domain, aligned-spin,
post-Newtonian point-particle approximant to model \ac{BNS}
injections~\citep{Buonanno:2009zt}, and a time-domain, effective-one-body waveform
calibrated against numerical relativity to model \ac{NSBH}
injections~\citep{Pan:2013rra,Taracchini:2013rva}. Waveform differences between these models and
the offline search templates are therefore including in the calculated $\langle VT \rangle$. 
The injected NSBH waveform model is not calibrated at high mass ratios
($m_1/m_2 >8$), so there is some additional modeling uncertainty for large-mass NSBH
systems. The true sensitive volume $\langle VT \rangle$  will also be smaller if the effect of
tides in \ac{BNS} or \ac{NSBH} mergers is extreme. However, for most scenarios
the effects of waveform modeling will be smaller than the
effects of calibration errors and the choice of prior discussed above. 

The posterior on $\Lambda$ (Eq.~\ref{eq:lambdapost}) can be reexpressed as a joint
posterior on the astrophysical rate $R$ and the sensitive volume  $\langle VT \rangle$
\begin{linenomath*}
\begin{equation}
p(R, \langle VT \rangle | s) \propto p(R, \langle VT \rangle) e^{-R \langle VT \rangle}.
\end{equation}
\end{linenomath*}
The new prior can be expanded as $p(R, \langle VT \rangle) = p(R | \langle VT \rangle) p(\langle VT \rangle)$. 
For $p(R | \langle VT \rangle)$, we will either use a uniform prior on $R$ or a prior proportional to the 
Jeffreys prior $1/\sqrt{R \langle VT \rangle}$. As with Refs.~\citep{Abbott:2016nhf, Abbott:2016drs, TheLIGOScientific:2016pea}, 
we use a log-normal prior on $\langle VT \rangle$
\begin{linenomath*}
\begin{equation}
p(\langle VT \rangle) = \ln \mathcal{N}(\mu, \sigma^2),
\end{equation}
\end{linenomath*}
where $\mu$ is the calculated value of $\ln\langle VT \rangle$ and $\sigma$ represents the fractional uncertainty
in $\langle VT \rangle$. Below, we will use an uncertainty of
$\sigma=18\%$ due mainly to calibration errors.

Finally, a posterior for the rate is obtained by marginalizing over $\langle VT \rangle$
\begin{linenomath*}
\begin{equation}
p(R | s) = \int d\langle VT \rangle\, p(R, \langle VT \rangle | s).
\label{eq:posterior}
\end{equation}
\end{linenomath*}
The upper limit $R_c$ on the rate with confidence $c$ is then given by the solution to
\begin{linenomath*}
\begin{equation}
\label{eq:upperlimit}
\int_0^{R_c} dR\, p(R | s) = c.
\end{equation}
\end{linenomath*}

For reference, we note that in the limit of zero uncertainty in $\langle VT \rangle$,
the uniform prior for $p(R | \langle VT \rangle)$ gives a rate upper limit of
\begin{linenomath*}
\begin{equation}
R_c = \frac{ -\ln(1-c) }{ \langle VT \rangle },
\end{equation}
\end{linenomath*}
corresponding to $R_{90\%} = 2.303/\langle VT \rangle$ for a 90\% confidence upper
limit~\citep{Biswas:2007ni}. For a Jeffreys prior on $p(R | \langle VT \rangle)$, this upper limit is
\begin{linenomath*}
\begin{equation}
R_c = \frac{ [{\rm erf}^{-1}(c)]^2 }{ \langle VT \rangle },
\end{equation}
\end{linenomath*}
corresponding to $R_{90\%} = 1.353/\langle VT \rangle$ for a 90\% confidence upper limit.

\subsection{BNS rate limits}
\label{ssec:bns_rate_limits}

\begin{figure}[t]
   \centering
   \includegraphics[width=\columnwidth]{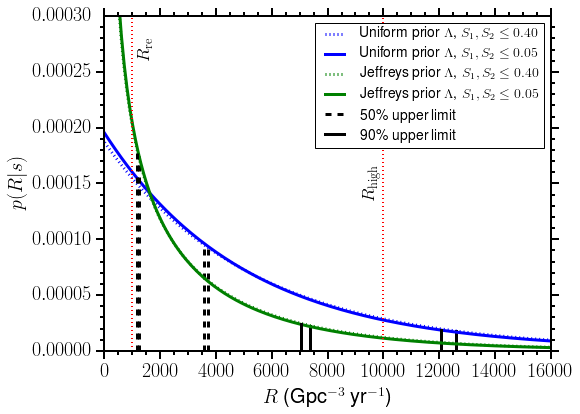} 
   \caption{Posterior density on the rate of \ac{BNS} mergers calculated using the \pycbc\ analysis.
   Blue curves represent
   a uniform prior on the Poisson parameter $\Lambda = R \langle VT \rangle$, while
   green curves represent a Jeffreys prior on $\Lambda$. The solid (low spin population)
   and dotted (high spin population) posteriors almost overlap. The vertical dashed and
   solid lines represent the 50\% and 90\% confidence upper limits respectively for each
   choice of prior on $\Lambda$. For each pair of vertical lines, the left line is the 
   upper limit for the low spin population and the right line is the upper limit for the high
   spin population. Also shown are the realistic $R_{\rm re}$ and high end
   $R_{\rm high}$ of the expected \ac{BNS} merger rates identified in Ref.~\citep{Abadie:2010cf}.}
   \label{fig:bnspdf}
\end{figure}

\begin{figure}[t]
\centering
\includegraphics[width=0.5\textwidth]{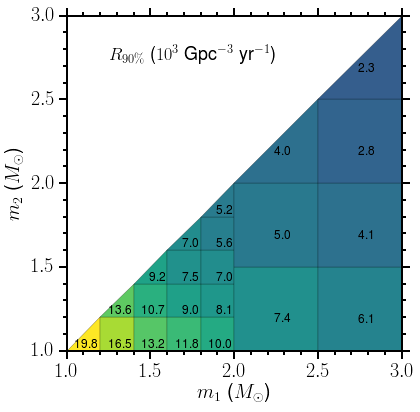}
\caption{\label{fig:bns_ul_vs_mass} 90\% confidence upper limit on the \ac{BNS} merger rate as a function of the two component
masses using the \pycbc\ analysis. Here the upper limit for each bin is obtained assuming a \ac{BNS} population with masses distributed uniformly
within the limits of each bin, considering isotropic spin direction and dimensionless spin magnitudes uniformly
distributed in $[0, 0.05]$.}
\end{figure}

Motivated by considerations in Section \ref{sec:source_considerations}, we begin by
considering a population of \ac{BNS} sources with a narrow range of component
masses sampled from the normal distribution $\mathcal{N}(1.35M_\odot, (0.13M_\odot)^2)$
and truncated to remove samples outside the range $[1,3] M_{\odot}$. We consider
both a ``low spin'' \ac{BNS} population, where spins are distributed with uniform dimensionless
spin magnitude $\in [0, 0.05]$ and isotropic direction, and a ``high spin'' \ac{BNS}
population with a uniform dimensionless spin magnitude $\in [0, 0.4]$ and isotropic direction.
Our population uses an isotropic distribution of sky location and source orientation and chooses
distances assuming a uniform distribution in volume.
These simulations are modeled using a post-Newtonian waveform model, expanded using the
``TaylorT4'' formalism~\citep{Buonanno:2009zt}.
From this population we compute the space-time volume that Advanced \ac{LIGO} was
sensitive to during the O1 observing run. Results are shown for the measured $\langle VT\rangle$
in Table~\ref{tab:bns_ul_table} using a detection threshold of $\mathrm{FAR} = 0.01\, \mathrm{yr}^{-1}$.
Because the template bank for the searches use only aligned-spin \ac{BNS} templates with 
component spins up to 0.05, the \pycbc\ (\gstlal) pipelines are 4\% (6\%) more sensitive 
to the low-spin population than to the high-spin population.
The difference in $\langle VT\rangle$ between the two analyses is no larger than 5\%, which
is consistent with the difference in time analyzed in the two analyses. 
In addition, the calculated $\langle VT \rangle$
has a Monte Carlo integration uncertainty of $\sim1.5\%$ due to the finite number of injection
samples.

\begin{table*}[t]
  \centering
  \begin{tabular}{c|c|c|c|c|c|c|c}
   Injection & Range of spin  & \multicolumn{2}{c|}{$\langle VT \rangle$ (Gpc$^3$~yr)} & \multicolumn{2}{c|}{Range (Mpc)} & \multicolumn{2}{c}{$R_{90\%}$ (Gpc$^{-3}$~yr$^{-1}$)} \\
   set & magnitudes & \pycbc & \gstlal & \pycbc & \gstlal & \pycbc & \gstlal \\
   \hline \hline
   Isotropic low spin & [0, 0.05] & \MainBNSVTPyCBCLowSpin & \MainBNSVTGstlalLowSpin & \MainBNSRangePyCBCLowSpin & \MainBNSRangeGstlalLowSpin & \MainBNSULPyCBCLowSpin & \MainBNSULGstlalLowSpin \\
   Isotropic high spin & [0, 0.4] & \MainBNSVTPyCBCHighSpin & \MainBNSVTGstlalHighSpin & \MainBNSRangePyCBCHighSpin & \MainBNSRangeGstlalHighSpin & \MainBNSULPyCBCHighSpin & \MainBNSULGstlalHighSpin \\
  \end{tabular}
  \caption{\label{tab:bns_ul_table} Sensitive space-time volume $\langle VT \rangle$ and 90\% confidence upper
  limit $R_{90\%}$ for \ac{BNS} systems. Component
  masses are sampled from a normal distribution $\mathcal{N}(1.35M_\odot, (0.13M_\odot)^2$) with samples outside the
  range $[1, 3]M_{\odot}$ removed. Values are shown for both the \texttt{pycbc}
  and \texttt{gstlal} pipelines. $\langle VT \rangle$ is calculated using a FAR threshold of 0.01~yr$^{-1}$. The 
  rate upper limit is calculated using a uniform prior on $\Lambda = R \langle
VT \rangle$ and an 18\% uncertainty 
  in $\langle VT \rangle$ from calibration errors.}
\end{table*}

Using the measured $\langle VT \rangle$, the rate posterior and upper limit can be
calculated from Eqs.~\ref{eq:posterior} and~\ref{eq:upperlimit} respectively.
The posterior and upper limits are shown in Figure~\ref{fig:bnspdf} and depend
sensitively on the choice of uniform versus Jeffreys prior
for $\Lambda=R\langle VT \rangle$. However, they depend only weakly on the spin
distribution of the \ac{BNS} population and on the width $\sigma$ of the uncertainty
in $\langle VT \rangle$. For the conservative uniform prior on $\Lambda$ and an
uncertainty in $\langle VT \rangle$ due to calibration errors of 18\%, we find the
90\% confidence upper limit on the rate of \ac{BNS} mergers to be \MainBNSULLowSpin~Gpc$^{-3}$~yr$^{-1}$
for low spin and \MainBNSULHighSpin~Gpc$^{-3}$~yr$^{-1}$ for high spin using the values of $\langle VT \rangle$
calculated with \pycbc; results for \gstlal\ are also shown in Table~\ref{tab:bns_ul_table}. These numbers can be compared to the upper limit 
computed from analysis of Initial \ac{LIGO} and Initial Virgo data~\citep{Colaboration:2011np}. 
There, the upper limit for $1.35$ -- $1.35 M_{\odot}$ non-spinning \ac{BNS} mergers is given 
as \SSixULNoSpin. The O1 upper limit is more than an order of magnitude lower than this previous upper limit.

To allow for uncertainties in the mass distribution of \ac{BNS} systems we also
derive 90\% confidence upper limits as a function of the \ac{NS} component masses.
To do this we construct a population of software injections with component masses
sampled uniformly in the range $[1, 3]M_{\odot}$, and an isotropic distribution
of component spins with magnitudes uniformly distributed in $[0, 0.05]$. We then bin
the \ac{BNS} injections by mass, and calculate $\langle VT \rangle$ and the associated 90\%
confidence rate upper limit for each bin. The 90\% rate upper limit for the conservative
uniform prior on $\Lambda$ as a function of component masses is shown in 
Figure~\ref{fig:bns_ul_vs_mass} for \pycbc. The fractional difference between the \pycbc\
and \gstlal\ results range from 1\% to 16\%.
%The upper limit for \gstlal\ is \red{$\sim x\%$} smaller.

\subsection{NSBH rate limits}
\label{ssec:nsbh_rate_limits}

\begin{table*}[t]
  \centering
  \begin{tabular}{c|c|c|c|c|c|c|c|c}
   NS mass & BH mass & Spin & \multicolumn{2}{c|}{$\langle VT \rangle$ (Gpc$^3$~yr)} & \multicolumn{2}{c|}{Range (Mpc)} & \multicolumn{2}{c}{$R_{90\%}$ (Gpc$^{-3}$~yr$^{-1}$)} \\
   ($M_\odot$) & ($M_\odot$) & distribution &  \pycbc & \gstlal & \pycbc & \gstlal & \pycbc & \gstlal \\
   \hline \hline
   1.4 & 5 & Isotropic & \MainNSBHVTPyCBCFiveIso\ & \MainNSBHVTGstlalFiveIso\ & \MainNSBHRangePyCBCFiveIso & \MainNSBHRangeGstlalFiveIso & \MainNSBHULPyCBCFiveIso\ & \MainNSBHULGstlalFiveIso\ \\
   1.4 & 5 & Aligned & \MainNSBHVTPyCBCFiveAligned\ & \MainNSBHVTGstlalFiveAligned\ & \MainNSBHRangePyCBCFiveAligned & \MainNSBHRangeGstlalFiveAligned & \MainNSBHULPyCBCFiveAligned\ & \MainNSBHULGstlalFiveAligned\ \\
   1.4 & 10 & Isotropic & \MainNSBHVTPyCBCTenIso\ & \MainNSBHVTGstlalTenIso\  & \MainNSBHRangePyCBCTenIso & \MainNSBHRangeGstlalTenIso & \MainNSBHULPyCBCTenIso\  & \MainNSBHULGstlalTenIso\ \\
   1.4 & 10 & Aligned & \MainNSBHVTPyCBCTenAligned\ & \MainNSBHVTGstlalTenAligned\ & \MainNSBHRangePyCBCTenAligned & \MainNSBHRangeGstlalTenAligned & \MainNSBHULPyCBCTenAligned\ & \MainNSBHULGstlalTenAligned\ \\
   1.4 & 30 & Isotropic & \MainNSBHVTPyCBCThirtyIso\ & \MainNSBHVTGstlalThirtyIso\ & \MainNSBHRangePyCBCThirtyIso & \MainNSBHRangeGstlalThirtyIso & \MainNSBHULPyCBCThirtyIso\ & \MainNSBHULGstlalThirtyIso\ \\
   1.4 & 30 & Aligned & \MainNSBHVTPyCBCThirtyAligned\ & \MainNSBHVTGstlalThirtyAligned\ & \MainNSBHRangePyCBCThirtyAligned & \MainNSBHRangeGstlalThirtyAligned & \MainNSBHULPyCBCThirtyAligned\ & \MainNSBHULGstlalThirtyAligned\ \\
  \end{tabular}
  \caption{\label{tab:nsbh_ul_table} Sensitive space-time volume $\langle VT \rangle$ and 90\% confidence upper
  limit $R_{90\%}$ for \ac{NSBH} systems with isotropic and aligned spin distributions. The NS spin magnitudes 
  are in the range $[0, 0.04]$ and the BH spin magnitudes are in the range $[0, 1]$. Values are shown for both the \texttt{pycbc}
  and \texttt{gstlal} pipelines. $\langle VT \rangle$ is calculated using a FAR threshold of 0.01~yr$^{-1}$. The 
  rate upper limit is calculated using a uniform prior on $\Lambda = R \langle
VT \rangle$ and an 18\% uncertainty 
  in $\langle VT \rangle$ from calibration errors.}
\end{table*}

\begin{figure}[t]
\centering
\includegraphics[width=0.5\textwidth]{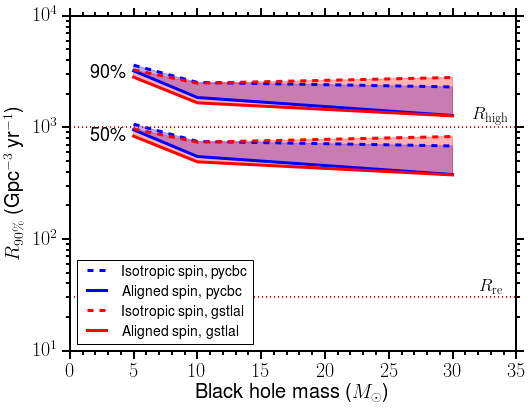}
\caption{\label{fig:nsbh_ul_vs_mass} 50\% and 90\% upper limits on the \ac{NSBH} merger rate
as a function of the \ac{BH} mass using the more conservative uniform prior for the counts $\Lambda$. 
Blue curves represent the \pycbc\ analysis and red curves
represent the \gstlal\ analysis. The \ac{NS} mass is assumed to be $1.4M_\odot$. The spin magnitudes
were sampled uniformly in the range [0, 0.04] for \acp{NS} and [0, 1] for \acp{BH}. For the aligned
spin injection set, the spins of both the \ac{NS} and \ac{BH} are aligned (or anti-aligned) with the
orbital angular momentum. For the isotropic spin injection set, the orientation for the
spins of both the \ac{NS} and \ac{BH} are sampled isotropically. The isotropic spin distribution results
in a larger upper limit. Also shown are the realistic $R_{\rm re}$ and high end
$R_{\rm high}$ of the expected \ac{NSBH} merger rates identified in Ref.~\citep{Abadie:2010cf}.}
\end{figure}

Given the absence of known \ac{NSBH} systems and uncertainty in the \ac{BH} mass, we evaluate the
rate upper limit for a range of \ac{BH} masses. We use three masses that span the likely
range of \ac{BH} masses: $5M_\odot$, $10M_\odot$, and $30M_\odot$. For the \ac{NS} mass,
we use the canonical value of $1.4M_\odot$. We assume a distribution of \ac{BH} spin magnitudes
uniform in $[0,1]$ and \ac{NS} spin magnitudes uniform in $[0, 0.04]$.
For these three mass pairs, we compute upper limits for an isotropic spin distribution
on both bodies, and for a case where both spins are aligned or anti-aligned with the orbital angular momentum
(with equal probability of aligned vs anti-aligned).
Our NSBH population uses an isotropic distribution of sky location and source orientation and chooses
distances assuming a uniform distribution in volume. Waveforms are modeled
using the spin-precessing, effective-one-body model calibrated against numerical relativity
waveforms described in Ref.~\citep{Taracchini:2013rva,Babak:2016tgq}.

The measured $\langle VT \rangle$ for a FAR threshold of $0.01 \mathrm{yr}^{-1}$ is given in Table~\ref{tab:nsbh_ul_table}
for \pycbc\ and \gstlal. The uncertainty in the Monte Carlo integration of $\langle VT \rangle$ is 1.5\%--2\%. The corresponding 
90\% confidence upper limits are also given using the conservative 
uniform prior on $\Lambda$ and an 18\% uncertainty in $\langle VT
\rangle$. Analysis-specific differences in the limits range from 1\% to 20\%,
comparable or less than other uncertainties such as calibration.  
These results can be compared to the upper limits found for initial \ac{LIGO} and Virgo
for a population of $1.35M_\odot$--$5M_\odot$ \ac{NSBH} binaries with isotropic spin of
\SSixNSBHULFiveSpin at 90\% confidence~\citep{Colaboration:2011np}.
As with the \ac{BNS} case, this is an improvement in the upper limit of over an order of magnitude.

We also plot the 50\% and 90\% confidence upper limits from \pycbc\ and \gstlal\ as a function of mass in 
Figure~\ref{fig:nsbh_ul_vs_mass} for the uniform prior. The search is
less sensitive to isotropic spins than to (anti-)aligned spins due to two factors.
First, the volume-averaged signal
power is larger for a population of (anti-)aligned spin systems than for isotropic-spin systems.
Second, the search uses a template bank of (anti-)aligned spin systems, and thus loses sensitivity
when searching for systems with significantly misaligned spins.
As a result, the rate upper limits are less constraining
for the isotropic spin distribution than for the (anti-)aligned spin case.